\documentclass[12pt, letterpaper]{article}

\pdfoutput=1
\usepackage[nottoc,notlot,notlof]{tocbibind}
\usepackage[title,titletoc]{appendix}

\usepackage{amsmath}
\usepackage{amsfonts}
\usepackage{amssymb}
\usepackage{graphicx, physics, subcaption}

\usepackage{color}

\usepackage{amsmath, amssymb, graphics, epsfig, graphicx}
\usepackage{epsf}
\usepackage{epstopdf}
\usepackage {amssymb}
\newcommand{\nc}{\newcommand}
\nc{\ba}{\begin{eqnarray}}
\nc{\ea}{\end{eqnarray}}
\newcommand\be{\begin{equation}}
\newcommand\ee{\end{equation}}

\newcommand{\bse}{\begin{subequations}}
\newcommand{\ese}{\end{subequations}}

\newcommand{\al}{\alpha}

\newcommand{\dg}{d^4{x}\sqrt{-g}}
\newcommand{\dge}{\dd[4]{x}\sqrt{g}}
\newcommand{\p}{\partial}

\begin{document}

\vspace{5mm}
\vspace{0.5cm}
\begin{center}

\def\thefootnote{\fnsymbol{footnote}}

{\bf\large Vacuum decay and bubble nucleation  in $f(R)$ gravity }
\\[0.5cm]

{ Borna Salehian$\footnote{salehian@ipm.ir },$
Hassan Firouzjahi$\footnote{firouz@ipm.ir }$
}
\\[0.5cm]

{\small \textit{School of Astronomy, Institute for Research in Fundamental Sciences (IPM) \\ P.~O.~Box 19395-5531, Tehran, Iran
}}\\

\end{center}

\vspace{.8cm}

\hrule \vspace{0.3cm}


\begin{abstract}

In this work we study vacuum decay and bubble nucleation in models of $f(R)$  higher curvature gravity. Building upon the analysis of Coleman-De Luccia (CDL), we present the formalism to  calculate the Euclidean action and the bounce solution for a  general $f(R)$ gravity  in the thin wall approximation. We calculate the size of the nucleated bubble  and the decay exponent for the Starobinsky model and its higher power extensions. We have shown that in the Starobinsky model  
with a typical potential  the nucleated bubble has a larger size in comparison to the CDL bubble  and with a lower tunneling rate. However, for higher power extension of the Starobinsky model the size of the bubble  and the tunneling exponent can be larger or smaller than the CDL bubble depending on the model parameters.  As a counterintuitive example, we have shown that a bubble with a larger size than the CDL bubble but  with a higher 
nucleation rate can be formed in $f(R)$ gravity.

\end{abstract}
\vspace{0.5cm} \hrule
\def\thefootnote{\arabic{footnote}}
\setcounter{footnote}{0}
\newpage

\section{Introduction}
\label{sec:intro}

Tunneling is a quantum mechanical phenomenon which does not occur in classical physics. 
 A minimum of the potential energy of the classical system is stable. However, in a quantum system where \(\hbar\) is not zero, uncertainty in the momentum causes the state of the system to become unstable. This is true for all minima of the potential energy, except the lowest one which remains stable. While the latter is called the ``true" vacuum the others are  called ``false" vacua.   

Vacuum tunneling plays important roles in various physical theories. For example, it is  important for the Standard Model of particle physics  since in the absence of new physics in higher energy scales, the vacuum of the Higgs boson can become unstable \cite{Sher:1988mj}. A similar but different issue is the 
 degenerate vacua of non-abelian gauge theories \cite{Belavin:1975fg}. In addition, it is believed that the landscape of string theory after compactification to four dimension posses many vacua. It is an ongoing debate how our primordial universe can emerge from tunneling in this vast landscape of vacua.  Our main interests here are focused on tunneling in early universe. Since the energy scale of early universe after big bang  is high new physics with multiple vacua can  play important roles.  Therefore, it is a well motivated question to study the effects of vacuum decay in early universe. In particular, tunneling and vacuum bubble nucleation played crucial roles in the development of old inflation scenarios  \cite{Guth:1980zm, Guth:1982pn}.

The primary question regarding vacuum decay is the probability of its occurrence which is normally addressed using semi-classical approximations.  The decay rate for a scalar field theory in flat spacetime 
was first studied by Coleman et. al. \cite{Coleman:1977py, Callan:1977pt}. It was then generalized to include the effects of gravity by Coleman and De Luccia  (CDL) \cite{Coleman:1980aw}, namely the ``CDL"  instanton (see also in \cite{Parke:1982pm}). It was shown that gravity can have significant effects on vacuum decay and bubble nucleation, specially for large bubbles, i.e. bubbles with  sizes comparable to the  horizon of the de Sitter background. 

One natural question then is to study vacuum decay and bubble nucleation in modified theories of gravity. 
In particular, the Starobinsky's model of inflation based on higher order curvature gravity \cite{Starobinsky:1980te} is one of the most successful scenario of inflation in early universe which is well consistent with the cosmological observations
\cite{Akrami:2018odb, Ade:2015lrj}. In addition, models of modified gravity are studied as candidates to explain the mysterious natures of dark matter or dark energy, for a review see \cite{Sotiriou:2008rp, DeFelice:2010aj}.  
Among different classes of modified gravity, here we consider \(f(R)\) theories in which the action is given by a general well-posed function of the Ricci scalar 
\be
S_G=\frac{1}{2}M_P^2\int\dg{}\,f(R)\,,
\ee
where \(M_P=1/{\sqrt{8\pi G}}\) is the reduced Planck mass. \(f(R)\) theories are simple but at the same time general enough to capture features of higher order curvature terms. To give some examples, Starobinsky's model of inflation is 
 given by \(f(R)=R+\al R^2\) with \(\al>0\) \cite{Starobinsky:1980te}. 
Theories with with \(f(R)=R-\al /R^n\) can cause late-time acceleration, though with drawbacks \cite{DeFelice:2010aj}. 

Related to our work, the  effects of Gauss-Bonnet term on tunneling is considered in \cite{Cai:2008ht,Charmousis:2008ce}. Also \cite{Czerwinska:2016fky} studied the effect of non-minimal coupling to gravity. More recently, tunneling in Jordan-Brans-Dickie theory was studied in \cite{Labrana:2018bkw}. 

The paper is organized as follows. In section \ref{sec:rev} we briefly review the formalism of vacuum decay  in flat spacetime. In section \ref{sec:fR} we present our analysis of  vacuum decay  in \(f(R)\) theory which is used in section \ref{sec:thinwall} for the thin wall limit. Section \ref{sec:example} is devoted to \(f(R)=R+\al R^n\) as an example of our setup. In section \ref{sec:einstein} we  reformulate the equivalent analysis  in the Einstein frame and finally we conclude in section \ref{sec:con}.

\section{ Vacuum decay   in flat spacetime }

\label{sec:rev}
\begin{figure}
		\centering
		\includegraphics[scale=0.5]{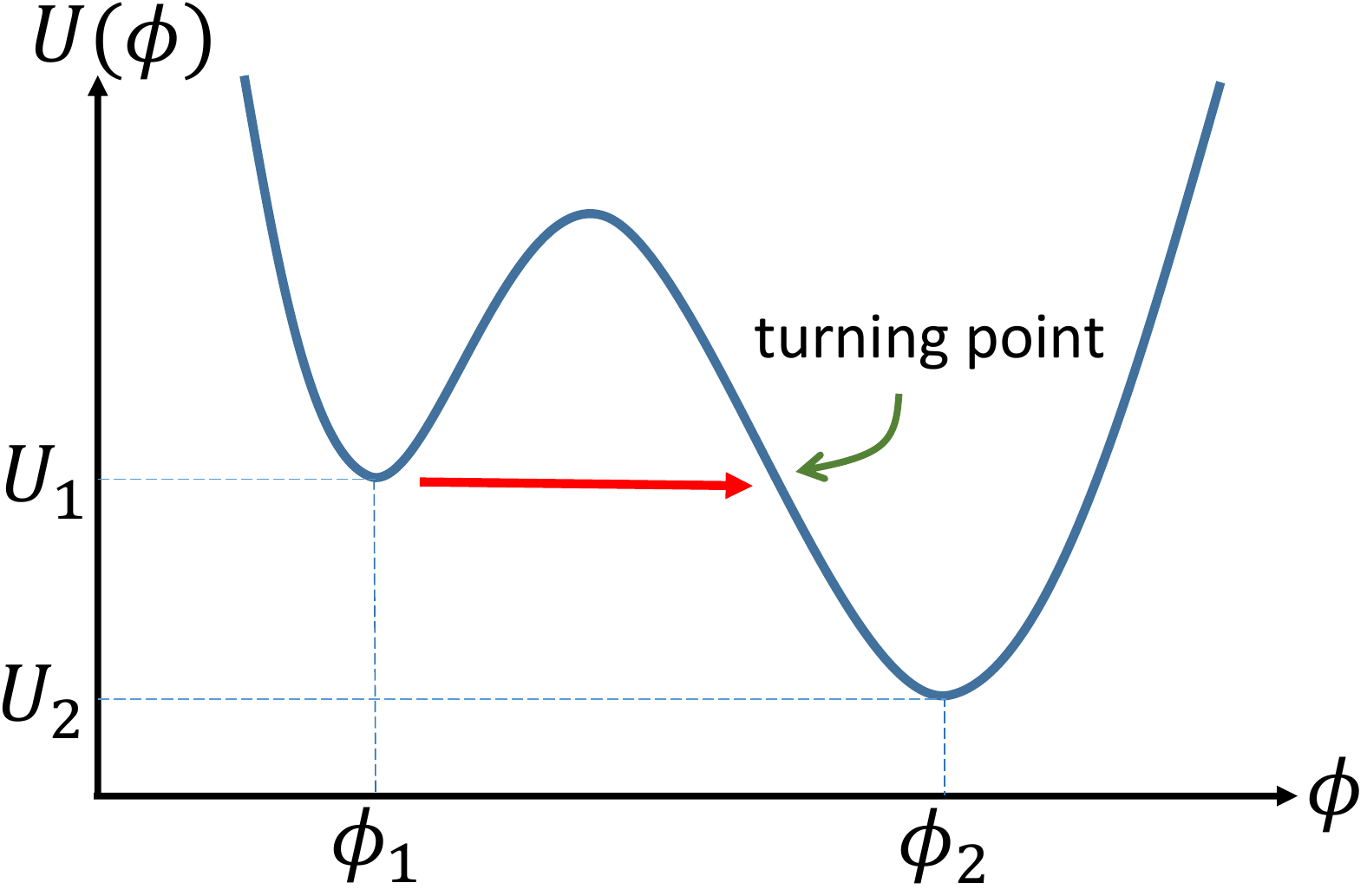}
		\label{fig:potential}	
	\caption{ The potential of the scalar field. False and true vacua are located respectively at \(\phi_1\) and \(\phi_2\). }
\label{fig:potconf}
\end{figure}

Here we briefly review the vacuum decay in flat spacetime. For an extensive review see  \cite{Weinberg:2012pjx} and
\cite{Coleman:1985rnk}. Consider the canonical scalar field
\be
\label{lm}
\mathcal{L}_m=-\frac{1}{2}(\partial_\mu\phi)^2-U(\phi)\,,
\ee
where the potential \(U(\phi)\) has two unequal minima as shown in Fig. \ref{fig:potconf}. We denote the field values at the 
false and true vacuum respectively by  \(\phi_1\) and \(\phi_2\) and the corresponding potential values  by  \(U_1\) and \(U_2\). Consider the situation where the field value is initially at \(\phi_1\) in all space. As mentioned above, there is a nonzero probability at every point that the field jumps to the turning point quantum mechanically. This is allowed as   tunneling respects energy conservation.

The conventional approach to compute the probability is to find the wave function of the system by solving the time-independent Schrodinger equation. The probability amplitude of tunneling is proportional to the ratio of the wave functions at the turning point and the false vacuum. For a field theory, the configuration space is infinite dimensional and we must solve for a wave functional. The wave functional can be found via WKB approximation under the potential barrier. However, the WKB equation is not easily solved for a multidimensional system even at zeroth order. The idea due to \cite{Banks:1973ps} is that the wave functional is maximized on a path in the configuration space, namely the most probable escape path (MPEP), and exponentially suppressed in its neighbourhood.  As a result, the multidimensional tunneling problem is reduced to a one dimensional case which is easily solved. 

It can be shown that there is a parameterization \(\tau\) of the MPEP curve in the configuration space that is a solution of the Euclidean equations of motion with appropriate boundary conditions \cite{Tanaka:1993ez,Bitar:1978vx}. This corresponds to replacing 
\(t\to -i\tau\) in the equation of motion, obtaining
\be
\label{eq:flatKG}
\frac{\partial^2}{\partial \tau^2}\phi+\laplacian{\phi}=\dv{U}{\phi}\,.
\ee
The boundary condition is that at \(\tau\to-\infty\) and \(\tau=0\), $\phi$ is at the initial and final configurations respectively. The tunneling probability amplitude can be shown to be related to the Euclidean action $S_E$ of this solution. Since we have spacetime translation symmetry, the probability per unit volume of spacetime is 
\be
\Gamma\propto\exp{-B}\,,
\ee
with the decay exponent \(B=S_E(\text{bounce})-S_E(\text{false vacuum})\). The Euclidean action is computed on a solution that passes over MPEP twice, the second time in reverse from the turning point to the false vacuum, hence the ``bounce" solution. It must then be subtracted from the false vacuum solution, i.e. in the absence of the bounce, 
where  \(\phi=\phi_1\) everywhere in spacetime.  The prefactor can be calculated using the path integral methods \cite{Callan:1977pt} but  its effect is subdominant compared to the exponential factor.

Coleman \cite{Coleman:1977py} assumed and then confirmed in \cite{Coleman:1977th} that the MPEP must be an \(O(4)\) symmetric solution of \eqref{eq:flatKG}. As a result, we must have
\be
\dv[2]{\phi}{\xi}+\frac{3}{\xi}\dv{\phi}{\xi}=\dv{U}{\phi}\,,
\ee
with \(\xi=\sqrt{\tau^2+\vb{x}^2}\) being the Euclidean distance from the origin. If we intuitively interpret \(\tau\) as the time,  the solution to this equation can be described as a growing bubble in which its interior is in true vacuum while 
its exterior is filled with  the  false vacuum. The bubble attains its maximum radius at \(\tau=0\) and then shrinks and disappears in \(\tau\to\infty\). 

If the difference between the false and true vacuum energy \(\epsilon\equiv U_1-U_2\) is small compared to the barrier height, it can be shown that the thickness of the wall can be neglected with respect to its radius. This is the thin wall approximation which can simplify the analysis considerably.  In the limit of thin wall approximation the energy of the wall can  be parameterized by the surface tension \(\sigma\) which can be calculated analytically.  In this limit, the size of the bubble at the time of nucleation is given by $\rho_0 = 3\sigma/\epsilon$. The tunneling exponent is then obtained to be 
\ba
\label{B0-eq}
B_0=\frac{\pi^2}{6}\rho_0^4\epsilon= \frac{27 \pi^2 \sigma^4}{2 \epsilon^3}  \, .
\ea
After formation, the bubble wall expands, approaching the speed of light asymptotically.     

This formalism was generalized by Coleman and De Luccia (CDL) \cite{Coleman:1980aw} to include the effects of Einstein gravity. It was shown that gravity can play important roles in vacuum tunneling and bubble nucleation, specially for large bubbles with sizes comparable to the Hubble radius of the background de Sitter spacetime.


\section{Euclidean action for $f(R)$ gravity}
\label{sec:fR}

In this section we present  our analysis of vacuum decay in $f(R)$ gravity in which the CDL solution can be obtained in the special case where $f(R) = R$. 

The starting  action is   
\be
S=\int\dg{}\left[\frac{1}{2}M_P^2f(R)+\mathcal{L}_m\right]+
S_{\text{bnd.}}\,,
\ee
in which the matter Lagrangian is the same as in Eq. (\ref{lm}) containing a scalar field with the false and true  vacua respectively at $\phi_1$ and $\phi_2$.  The boundary term $S_{\text{bnd.}}$ is added to take care of the total derivative terms. 
In Einstein theory of general relativity (GR)  this is done by adding a total divergence term, the Gibbons-Hawking-York term. There is not a unique similar way in \(f(R)\) theories \cite{Sotiriou:2008rp}. However, in correspondence with scalar-tensor theories one can obtain \cite{Dyer:2008hb}
\be
S_{\text{bnd.}}=M_P^2\int\dd[3]x\sqrt{h}\;F(R)K\,,
\ee
where \(h\) is the determinant of the induced metric on the boundary, \(K\) is trace of the extrinsic curvature and \(F(R)\equiv \dv{f}{R}\). 
	
The equations of motion  are obtained by varying the action with respect to metric, yielding \cite{DeFelice:2010aj}
\be
\label{eq:eins}
\Sigma_{\mu\nu}=8\pi G\:T_{\mu\nu}\,,
\ee
in which  
\be
\label{eq:sigmamunu}
\Sigma_{\mu\nu}=F(R)G_{\mu\nu}+\frac{1}{2}(RF(R)-f(R))g_{\mu\nu}-[\nabla_{\mu}\nabla_{\nu}-g_{\mu\nu}\Box]F(R)\,,
\ee
is the generalization of the Einstein tensor and 
\be
T_{\mu\nu}=\frac{-2}{\sqrt{-g}}\fdv{(\sqrt{-g}\mathcal{L}_m)}{g^{\mu\nu}}\,,
\ee
is the energy momentum tensor.  For the scalar field  we have
\be
T_{\mu\nu}=\p_\mu\phi\p_\nu\phi+g_{\mu\nu}\mathcal{L}_m\,.
\ee
Another useful equation is the trace of \eqref{eq:eins} which reads
\be
\label{eq:tr}
RF(R)-2f(R)+3\Box{F(R)}=8\pi G\:T\,,
\ee
where \(T\) is the trace of the energy momentum tensor. 

The Klein-Gordon equation also reads
\be
\label{eq:kg}
\frac{1}{\sqrt{-g}}\p_\mu\left(\sqrt{-g}\:\p^\mu\phi\right)=\dv{U}{\phi}\,.
\ee
	
As discussed above, we have to write the Euclidean action with positive definite signature metric. This is given by
\be
\label{eq:se}
S_E=\int\dge{}\left[-\frac{1}{2}M_P^2f(R)+\mathcal{L}^E_m\right]-S_{\text{bnd.}}\,,
\ee
where 
\be
\mathcal{L}^E_m=\frac{1}{2}(\partial_\mu\phi)^2+U(\phi)\,.
\ee
If we define the Euclidean energy momentum tensor
\be
T^E_{\mu\nu}  =\frac{2}{\sqrt{g}}\fdv{(\sqrt{g}\mathcal{L}^E_m)}{g^{\mu\nu}}\\
=\p_\mu\phi\p_\nu\phi-g_{\mu\nu}\mathcal{L}^E_m\,,
\ee
then Eqs. \eqref{eq:eins} and \eqref{eq:tr} are still valid with everything computed with the Euclidean metric. From now on we drop the index \(E\) and, otherwise stated,  work with the Euclidean equations.

As in the case of CDL, we assume  that the tunneling solution is dominated by the \(O(4)\) symmetric solutions of the Euclidean equations of motion. The most general form of the line element with this symmetry is 
\be
\label{eq:ds2}
\dd{s}^2 = \dd{\xi}^2+\rho(\xi)^2\dd{\Omega}^2_3\\
     = \dd{\xi}^2+\rho(\xi)^2\left(\dd{\chi}^2+\cos^2\chi\dd{\Omega}^2\right)\,,
\ee
where \(\dd{\Omega}^2_3\) and \(\dd{\Omega}^2\) are the line elements of \(S_3\) and \(S_2\) respectively. The curvature of the \(\xi=\text{const.}\) three sphere is given by \(\frac{6}{\rho(\xi)^2}\) so $\rho(\xi)$  represents the radius of the sphere in curved spacetime. In comparison, in flat spacetime $\rho(\xi) = \xi$.  

The field \(\phi\) is only a function of \(\xi\) and the equation becomes
\be
\phi''+\left(\frac{3\rho'}{\rho}\right)\phi'=\dv{U}{\phi}\,,
\ee
where a prime means differentiation with respect to \(\xi\). 

The angular part of the integration in the Euclidean action \eqref{eq:se} is trivial and we obtain
\be
\label{eq:se_inter}
S_E=2\pi^2\int\dd{\xi}\rho^3\left[-\frac{1}{2}M_P^2f(R)+\frac{1}{2}{\phi'}^2+U\right]-S_{\text{bnd.}}\,,
\ee
where the Ricci scalar is given by
\be
\label{eq:ricci}
R=-\frac{6}{\rho^2}(\rho'^2+\rho\rho''-1)\,.
\ee
This can be simplified by eliminating the kinetic term with the help of equations of motion
\be
M_P^2\Sigma_{\xi\xi}=T_{\xi\xi}\\
=\frac{1}{2}{\phi'}^2-U\,.
\ee
On the other hand, from \eqref{eq:sigmamunu} we have
\be
\Sigma_{\xi\xi}=\left(\frac{-3\rho''}{\rho}\right)F-\frac{1}{2}f(R)+\left(\frac{3\rho'}{\rho}\right)F'\, .
\ee
Using these equations, the action is simplified to 
\be
S_E=2\pi^2M_P^2\int\dd{\xi}\left\{\rho^3\left[\frac{2U}{M_P^2}-f(R)\right]-3\rho^2\rho''F+3\rho^2\rho'F'\right\}-S_{\text{bnd.}}\,.
\ee
If we do the integration by part on the last term in the braces we obtain
\be
S_E=2\pi^2M_P^2\int\dd{\xi}\left\{\rho^3\left[\frac{2U}{M_P^2}-f(R)\right]-6\rho ({\rho'}^2+\rho\rho'') F \right\}+2\pi^2M_P^2\rho^3\left(\frac{3\rho'}{\rho}\right) F \Bigg|_b-S_{\text{bnd.}}\,.
\ee
The surface term can be shown to be canceled by the boundary term since the trace of the extrinsic curvature of 
the hypersurface of  \(\xi=\text{const.}\)  is \(K=\frac{3\rho'}{\rho}\). In addition, the second term in the integral can be rewritten using the expression of the Ricci scalar Eq. \eqref{eq:ricci} and we finally obtain
\be
\label{eq:se_final}
S_E=4\pi^2M_P^2\int\dd{\xi}\left\{\rho^3\left[\frac{U}{M_P^2}+\frac{1}{2}(RF(R)-f(R))\right]-3\rho F(R)\right\}\,.
\ee
For the Einstein GR the above action  reduces to
\be
S_E=4\pi^2M_P^2\int\dd{\xi}\left[\rho^3\frac{U}{M_P^2}-3\rho\right]\,,
\ee
which is the expression used originally in \cite{Coleman:1980aw}. 

We need to compute the Euclidean action for the false vacuum and the bounce solution. In the remaining of this section we perform the calculations for the false vacuum solution and devote the next section to the analysis of bounce solution. 

We are looking for the solution of the equations of motion when the field is sitting at its false vacuum where \(\phi(\xi)=\phi_1\). Since by assumption the symmetry group of the solution is \(O(4)\), we have \(6(=\frac{4\times3}{2})\) Killing vectors for different rotations at each point. Additionally, for the case when the field is constant over the spacetime we also have \(4\) more Killing vectors for spacetime translation invariance. So, we have \(10\) Killing vectors overall which is the maximum number of symmetry transformations for a four-dimensional  spacetime,  i.e.  the spacetime is maximally symmetric.
	
It can be shown that for any maximally symmetric spacetime the Ricci scalar is a constant. If we look at the trace of the generalized Einstein field equation,  Eq.  \eqref{eq:tr},  we get
\be
\label{eq:desric}
RF(R)-2f(R)=8\pi G\:T\,,
\ee
where \(T=-4U_1\). 

Let us introduce \(\Lambda_1^2 \equiv \frac{3M_P^2}{U_1}\) as the length scale associated with the potential where we have assumed that \(U_1\) is positive. Equation (\ref{eq:desric}) is an algebraic equation for the Ricci scalar. The solution to this equation, whatever it is, we call it \(R_1\equiv \frac{12}{L_1^2}\) to account for the dimension of the curvature. Thus \(L_1\) is the length scale that characterizes the curvature of the spacetime.  It is important to note that for a general $f(R)$ theory $L_1 \neq \Lambda_1$. However, one can check that for the Einstein GR, they are equal. 

Eq. \eqref{eq:desric} may not have  any positive solution for some theories or even it may contain more than one solution. However, here we assume that we have one and only one positive solution for both false and true vacuum. 
	
The function \(\rho(\xi)\) can be deduced using the \(\xi\xi\) component of the generalized Einstein equation,  Eq. (\ref{eq:eins}) 
\be
\label{eq:xixi}
\frac{3}{\rho^2}({\rho'}^2-1)F_1+\frac{1}{2}(R_1F_1-f_1)=\frac{-3}{\Lambda_1^2}\,,
\ee
where \(F_1\equiv F(\frac{12}{L_1^2})\) and \(f_1 \equiv f(\frac{12}{L_1^2})\). 

Combining the above equation with Eq. \eqref{eq:desric} we obtain
\be
{\rho'}^2=1-\frac{\rho^2}{L_1^2}\,.
\ee
The most general solution to this equation is 
\be
\rho(\xi)=L_1\sin{\left(\frac{\xi}{L_1}\right)}\,,
\ee
in which a constant of integration is eliminated by an appropriate shift of coordinate such that \(\rho(\xi=0)=0\). 
This is the well known expression for de Sitter solution in GR. Note that in our $f(R)$ setup 
 $L_1$, and not $\Lambda_1$, appears in $\rho(\xi)$. This is understandable since $L_1$ determines  the curvature scale of the spacetime.

Similar expressions are also valid if \(U_1\) is zero or negative. Let us assume that Eq. \eqref{eq:desric} has Minkowski solution with \(R_1= U_1 =0\) or anti de Sitter (AdS) solution with \(R_1=-\frac{12}{L_1^2}\),  corresponding to   \(U_1=-\frac{3M_P^2}{\Lambda_1^2}\). The corresponding solution then becomes, 
\be
\rho(\xi)=
\begin{cases}
L_1\sin{\left(\frac{\xi}{L_1}\right)}&\:\text{dS}\\
\xi&\:\text{Minkowski}\\
L_1\sinh{\left(\frac{\xi}{L_1}\right)}&\:\text{AdS}\\
\end{cases}\,.
\ee
Note that the spacetime is closed with no boundary for dS and the integration in Eq. \eqref{eq:se_final} is over \([0,\pi L_1]\) which is bounded by the zeros of \(\rho(\xi)\). However, for the other cases, the spacetime is open with boundary at infinity and the integration runs over \([0,\infty)\). Finally, we can obtain Minkowski and AdS solutions from dS by \(L_1\to\infty\) and \(L_1\to iL_1\). Thus, in what follows,  we assume the potential is positive everywhere and the tunneling is from dS to dS spacetimes.  

\section{Decay exponent in the thin wall limit}
\label{sec:thinwall}

To compute the decay rate, we need the Euclidean action for the bounce solution. This is the nontrivial solution of the equations of motion Eqs. \eqref{eq:tr} and \eqref{eq:kg} with the boundary conditions which for the closed spacetime are
\be
\phi'(0)=\phi'(\xi_m)=0\,,
\ee
where \(\xi_m\) is the second zero of the function \(\rho(\xi)\). 

\begin{figure}
	\centering
		\includegraphics[scale=0.5]{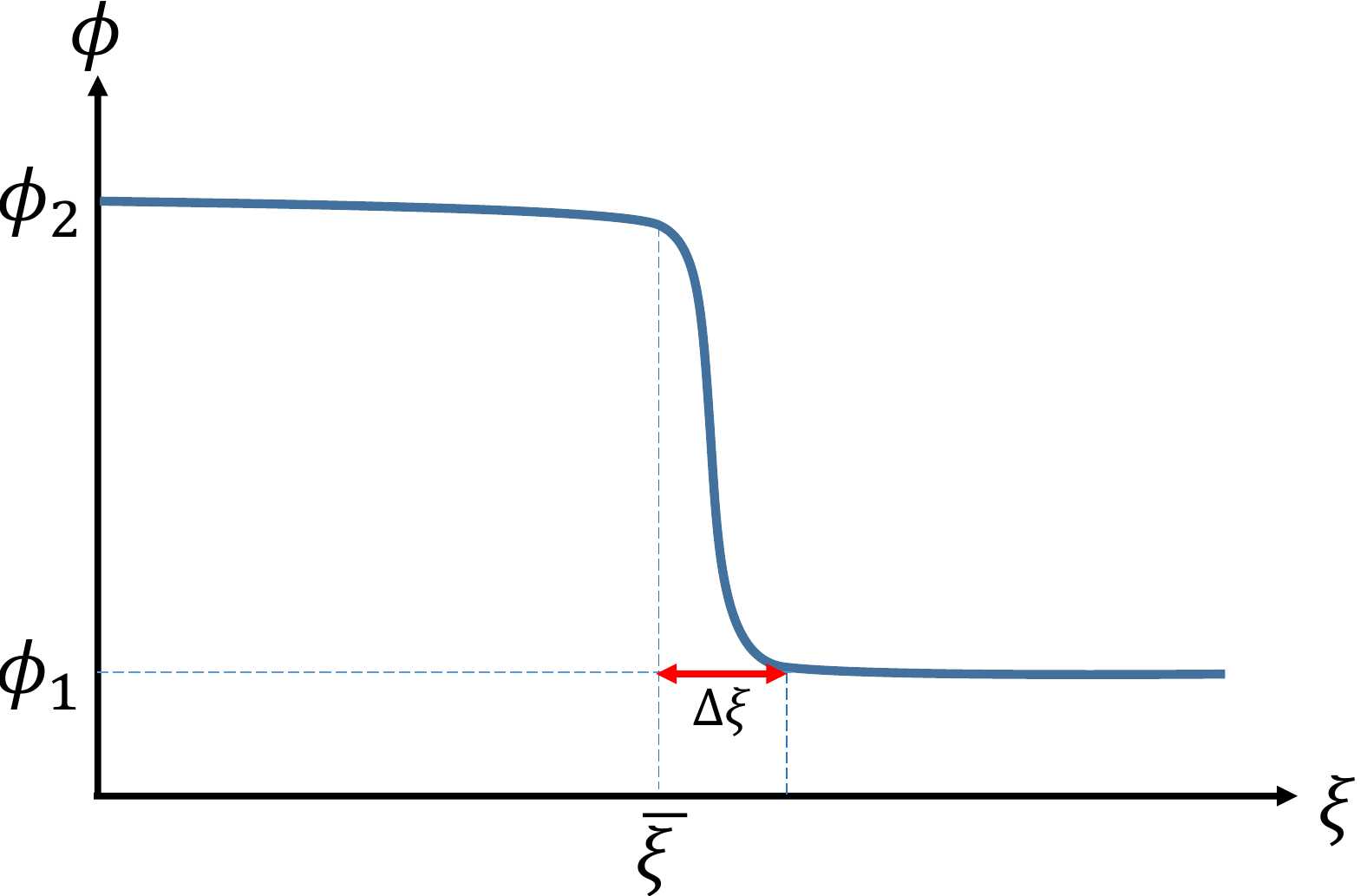}
		\includegraphics[scale=0.5]{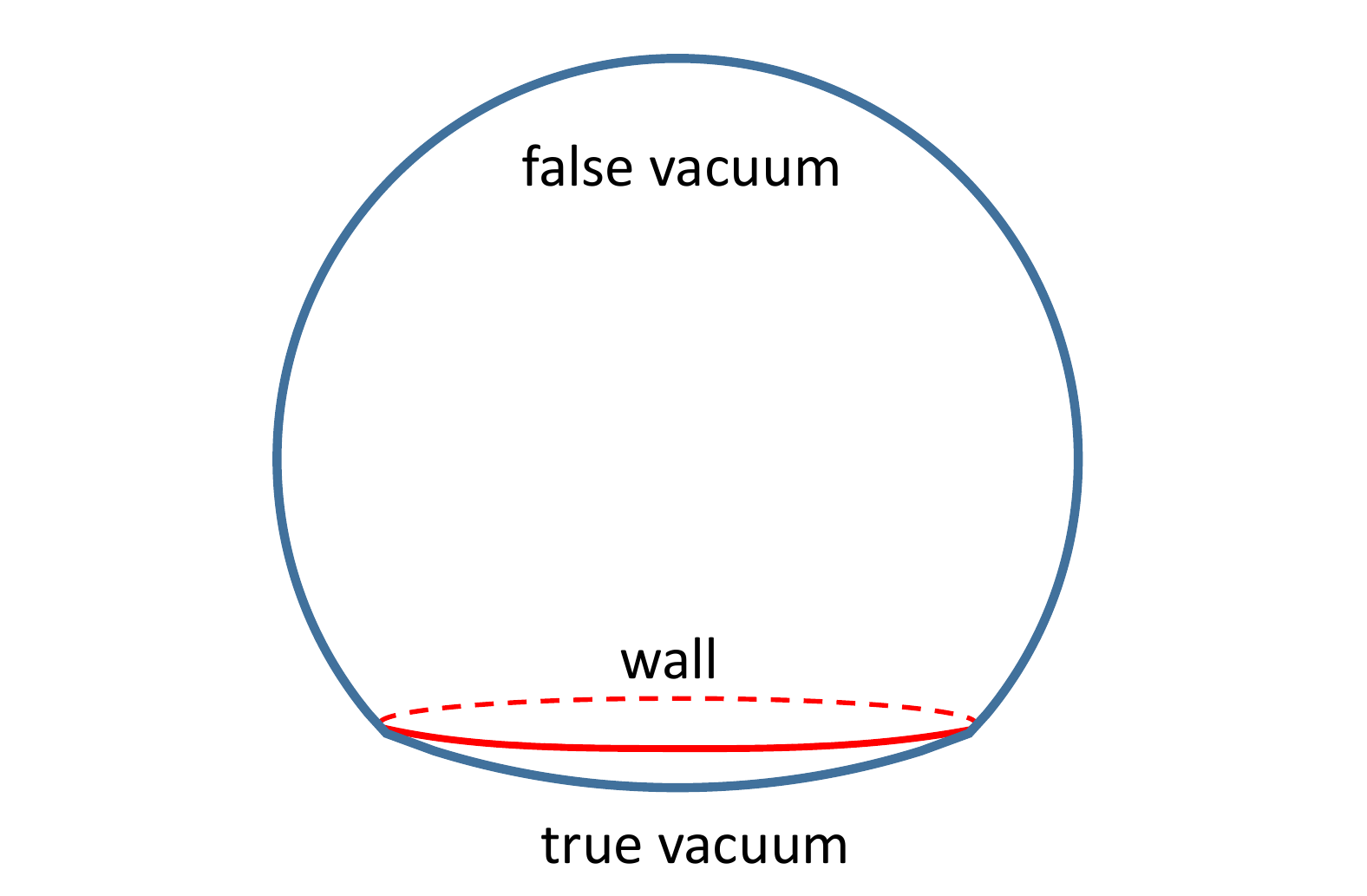}
	\caption{Left: The solution of the scalar field in the thin wall limit in which the transition occurs in a  narrow region. Right: The structure of the type A bounce solution where a dS space is viewed as an 
	embedding of a  four-sphere inside a five dimensional Euclidean spacetime. As shown, the bounce solution consists two  four-spheres with different radii. The position of the wall is further from the equator of the false vacuum.}
\label{fig:phibubble}
\end{figure}

It is not easy to find such a solution for a general potential, nor is it to prove its existence. In analogy to the CDL analysis, we assume that an instanton solution exists in which the variation of \(\phi\) occurs in a small region near the bubble wall. We also assume that everywhere inside the bubble is filled with true vacuum while outside the bubble the state is in false vacuum. This is illustrated in Fig. \ref{fig:phibubble}.

Quantitatively, in the thin wall limit we assume \(\Bar{\rho}\gg\rho'\Delta\xi\), where \(\Delta\xi\) is the wall thickness and \(\Bar{\rho}=\rho(\Bar{\xi})\) is the bubble radius with $\bar \xi$ being the coordinate position of the wall.
As a result, we can find \(\phi(\xi)\) inside the wall approximately. More specifically,  Eq. \eqref{eq:kg} can be written as
\be
\label{eq:kgwall}
\frac{1}{2}{\phi'}^2-U(\phi)=-U_1+\int_{\bar{\xi}+\Delta\xi}^\xi\dd{\xi}\frac{3\rho'}{\rho}{\phi'}^2\,.
\ee
However, in the thin wall limit, the integral over the friction term above can be discarded and we  approximately have
\be
\label{eq:kg2}
\frac{1}{2}{\phi'}^2\approx U(\phi)-U_1\,. 
\ee
For the later use,  this yields 
\be
\label{eq:dxi}
\dd{\xi}=\frac{\dd{\phi}}{\sqrt{2(U(\phi)-U_1)}}\,.
\ee
The function \(\rho(\xi)\) is approximately known everywhere
\be
\rho(\xi)=
\begin{cases}
L_2\sin{\left(\frac{\xi}{L_2}\right)}&\:0\leq\xi\leq\Bar{\xi}\\
\rho_{\text{wall}}(\xi)&\:\Bar{\xi}\leq\xi\leq\Bar{\xi}+\Delta\xi\\
L_1\sin{\left(\frac{\xi_m-\xi}{L_1}\right)}&\:\Bar{\xi}+\Delta\xi\leq\xi\leq\xi_m\\
\end{cases}\,,
\ee
where $L_2$ is defined (similar to $L_1$) as the solution of Eq. \eqref{eq:desric} for the true vacuum with $T= - 4 U_2$. 
We have assumed  \(\rho_{\text{wall}}(\xi)\approx\bar{\rho}\) since we neglect the variation of \(\rho\) inside the wall. 

The form of the solution is such that the conditions \(\rho(0)=\rho(\xi_m)=0\) are trivially satisfied. By the continuity of the metric we must have
\be
\label{eq:controb}
\bar{\rho}=L_2\sin{\left(\frac{\bar{\xi}}{L_2}\right)}=L_1\sin{\left(\frac{\xi_m-\Bar{\xi}-\Delta\xi}{L_1}\right)}\,,
\ee
thus from the three constants \(\bar{\rho}\), \(\bar{\xi}\) and \(\xi_m\) only one is independent (neglecting the wall thickness). Here, we assume for the arguments 
\be
\label{eq:typeA}
0< \frac{\bar{\xi}}{L_2}<\frac{\pi}{2} , \qquad \qquad 
\frac{\pi}{2}< \frac{\xi_m-\Bar{\xi}-\Delta\xi}{L_1}<\pi \, .
\ee
The meaning of these conditions become more evident when we consider embedding of a dS spacetime as a sphere inside a five dimensional Euclidean spacetime. As a result, the bounce solution is a composite of two spheres with different radii (see Fig. \ref{fig:phibubble}). Conditions \eqref{eq:typeA} mean that the position of the wall is further from the hemisphere in the false vacuum region. This is reasonable since we expect that the size of the bubble at the moment of formation to be
much smaller than the size of the dS horizon. In the classification of \cite{Weinberg:2012pjx,Marvel:2007pr} this is the type A bounce solution.  Note that only the type A bounce solution exists in the flat spacetime solution in the limit of weak gravity. Therefore, in this work, we consider only the type A bounce solution.

In GR, solving \(\phi(\xi)\) fixes the Ricci scalar algebraically  through the trace of the Einstein equations. However, here we need to solve a differential equation 
\be
\label{eq:wall}
3\Box{F(R)}+RF(R)-2f(R)=-\frac{1}{M_P^2}(6U(\phi)-2U_1)\,,
\ee
where we have used Eq. \eqref{eq:kg2} to eliminate the the kinetic term. It may not be easy to solve this equation in general.  

The decay exponent is the Euclidean action for the  bounce solution minus the Euclidean action for the 
false vacuum solution. The Euclidean action is given in Eq. (\ref{eq:se_final}).  We can split the integral in 
Eq. (\ref{eq:se_final}) into three regions, inside the wall, on the wall and outside the wall, both for the bounce and the false vacuum solution, 
\ba
B_{\text{in}}&=&S_E(\text{b})\bigg|_0^{\bar{\xi}}-S_E(\text{fv})\bigg|_0^{\bar{\xi}}\\
B_{\text{w}}&=&S_E(\text{b})\bigg|_{\bar{\xi}}^{\bar{\xi}+\Delta\xi}-S_E(\text{fv})\bigg|_{\bar{\xi}}^{\bar{\xi}+\Delta\xi}\\
B_{\text{out}}&=&S_E(\text{b})\bigg|_{\bar{\xi}+\Delta\xi}^{\xi_m}-S_E(\text{fv})\bigg|_{\bar{\xi}+\Delta\xi}^{\pi L_1}\,,
\ea
where the vertical lines show the integration limits.  Below we calculate each contributions separately. 

Since outside the wall the bounce solution is at false vacuum, by a change of the variable of the integral, we can write
\be
B_{\text{out}}=S_E(\text{fv})\bigg|_{\pi L_1}^{\xi_m}\,.
\ee
Note that if the spacetime is open as in  Minkowski or AdS cases this is zero since the integral is up to infinity. 
For dS case, from Eqs.  \eqref{eq:controb} and \eqref{eq:typeA},  we can write 
\be
\xi_m-\pi L_1=L_2\arcsin{\frac{\bar{\rho}}{L_2}}-L_1\arcsin{\frac{\bar{\rho}}{L_1}}+\Delta\xi\,,
\ee
so the integration domain is of \(\order{\epsilon}\) hence we may neglect \(B_{\text{out}}\). 

For \(B_{\text{in}}\),  from \eqref{eq:se_final},  we get
\ba
\label{eq:Bin}
B_{\text{in}}&=&4\pi^2M_P^2\int_0^{\bar{\xi}}\dd{\xi}\Big\{\rho^3 \Big[\frac{3}{\Lambda_2^2}+\frac{1}{2}(R_2F_2-f_2)\Big]-3\rho F_2\Big\}-\Big\{\rho^3\Big[\frac{3}{\Lambda_1^2}+\frac{1}{2}(R_1F_1-f_1)\Big]-3\rho F_1\Big\} \nonumber \\
&=&12\pi^2M_P^2\Big(F_2\int_0^{\bar{\xi}}\dd{\xi}\Big[\frac{\rho^3}{L_2^2}-\rho\Big]-F_1\int_0^{\bar{\xi}}\dd{\xi}\Big[\frac{\rho^3}{L_1^2}-\rho\Big]\Big) \nonumber \\
&=&4\pi^2M_P^2F_2L_2^2\Big[\Big(1-\frac{\Bar{\rho}^2}{L_2^2}\Big)^{3/2}-1\Big]-4\pi^2M_P^2F_1L_1^2\Big[\Big(1-\frac{\Bar{\rho}^2}{L_1^2}\Big)^{3/2}-1\Big]\,,
\ea
where we have used Eq. \eqref{eq:xixi} in the second line. 

Finally, the contribution from the wall is 
\ba
B_{\text{w}} &=&2\pi^2\bar{\rho}^3\int_{\text{wall}}\dd{\xi}\Big [ 2 \Big(U(\phi)-U_1 \Big)+M_P^2 \Big(RF(R)-f(R) \Big)-M_P^2 \Big(R_1F(R_1)-f(R_1) \Big)\Big] \nonumber\\
\quad&-&12\pi^2M_P^2\bar{\rho}\int_{\text{wall}}\dd{\xi}\Big[F(R)-F(R_1)\Big]\,,
\label{eq:bwall1}
\ea 
where the integration is over the wall i.e. \(\bar{\xi}<\xi<\bar{\xi}+\Delta\xi\). The first term under the first integral, using Eq. \eqref{eq:dxi}, can be written  as
\ba
\sigma &\equiv &\int_{\text{wall}}\dd{\xi}2(U(\phi)-U_1) \nonumber \\
&=&\int_{\text{wall}}\dd{\phi}\sqrt{2(U(\phi)-U_1)} \, ,
\ea	
which is the surface tension of the bubble in the absence of gravity. Note that in GR, this is also the only surviving term in \eqref{eq:bwall1}. However, in a general \(f(R)\) theory gravity has nonzero contributions to $B_{\text{w}}$ even in the thin wall limit. 

To put Eq. \eqref{eq:bwall1} into a more useful form, let us use Eq. \eqref{eq:tr} to write
\be
M_P^2(RF(R)-f(R))=T+M_P^2f(R)-3M_P^2\Box F(R)\,.
\ee
Since everything is only a function of \(\xi\), we have
\be
\Box F(R)=F''+\frac{3\rho'}{\rho}F'\,.
\ee
The integration of the first term can be done and gives \(F'\Big|_2^1\) which is zero since the Ricci scalar and any function of it are constant in either sides of the wall. The integral of the second term can also be neglected, by the same argument that we have neglected the friction term in \eqref{eq:kgwall}. Finally, using Eq. \eqref{eq:kg2}, 
the trace of the energy-momentum tensor can also be written in terms of the potential,   so at the end we obtain 
\ba
\label{Bw}
B_{\text{w}}&\equiv&2\pi^2\bar{\rho}^3S_1-12\pi^2\bar{\rho}S_2 \nonumber\\
&\equiv&2\pi^2\bar{\rho}^3\left\{\int_{\text{wall}}\dd{\xi}M_P^2(f(R)-f_1)-2\sigma\right\}-12\pi^2\bar{\rho}\int_{\text{wall}}\dd{\xi}M_P^2\big(F(R)-F_1 \big)\,,
\ea   
where we have introduced \(S_1\) and \(S_2\) to account for the new terms. For GR, the integral in the first bracket is \(3\sigma\) and the second integral vanishes i.e. \(S_1=3\sigma - 2 \sigma=\sigma\) and \(S_2=0\) and everything is consistent. 

Combining the contribution from $B_{\text{w}}$ and $B_{\text{in}}$ from Eqs. (\ref{eq:Bin}) and (\ref{Bw}),  the final expression for the decay exponent can be written as 
\ba
\label{eq:B}
B=2\pi^2 \Bigg[
\Bar{\rho}^3S_1-6\bar{\rho}S_2-2M_P^2F_1L_1^2\Big(\big(1-\frac{\Bar{\rho}^2}{L_1^2}\big)^{3/2}-1\Big)+2M_P^2F_2L_2^2\Big(\big(1-\frac{\Bar{\rho}^2}{L_2^2}\big)^{3/2}-1\Big) 
\Bigg] .
\ea
In order to find \(\Bar{\rho}\), the size of the bubble at the time of nucleation,  
we need to extremize the decay exponent given in Eq. \eqref{eq:B}, 
\ba
\label{eq:Bderiv}
\dv{B}{\Bar{\rho}}&=&12\pi^2M_P^2F_1\Bar{\rho}\sqrt{1-\frac{\Bar{\rho}^2}{L_1^2}}-12\pi^2M_P^2F_2\Bar{\rho}\sqrt{1-\frac{\Bar{\rho}^2}{L_2^2}}+6\pi^2\Bar{\rho}^2S_1-12\pi^2S_2 \nonumber\\
&=&12\pi^2M_P^2\left\{F_1\Bar{\rho}\sqrt{1-\frac{\Bar{\rho}^2}{L_1^2}}-F_2\Bar{\rho}\sqrt{1-\frac{\Bar{\rho}^2}{L_2^2}}+s_1\Bar{\rho}^2-s_2\right\}=0\,,
\ea
where we have introduced lowercase \(s_i\) defined via 
\ba
s_1\equiv \frac{S_1}{2M_P^2} \, ,  \quad \quad 
s_2 \equiv \frac{S_2}{M_P^2}\,.
\ea
Upon twice squaring, and defining \(x\equiv \bar{\rho}^2\), the  equation can be written as 
\be
\label{eq:z}
C_1x^4+C_2x^3+C_3x^2+C_4x+C_5=0\,,
\ee
where
\begin{align}
C_1&=\left(s_1^2+\frac{F_1^2}{L_1^2}+\frac{F_2^2}{L_2^2}\right)^2-4\frac{F_1^2}{L_1^2}\frac{F_2^2}{L_2^2} \nonumber\\
C_2&=4F_1^2F_2^2\left(\frac{1}{L_1^2}+\frac{1}{L_2^2}\right)-2(2s_1s_2+F_1^2+F_2^2)\left(s_1^2+\frac{F_1^2}{L_1^2}+\frac{F_2^2}{L_2^2}\right) \nonumber\\
C_3&=(2s_1s_2+F_1^2+F_2^2)^2-4F_1^2F_2^2 \nonumber\\
C_4&=-2s_2^2(2s_1s_2+F_1^2+F_2^2) \nonumber\\
C_5&=s_2^4\,.  \nonumber
\end{align}

 In general, the solution of  Eq. \eqref{eq:z} is  given by 
\be
x=\bar{\rho}^2 = -\frac{C_2}{4C_1}-\frac{1}{2}\sqrt{I}-\frac{1}{2}\sqrt{II}\,,
\ee
where \(I\) and \(II\) are some complicated functions of the coefficients. The decay exponent is then computed if we substitute this solution into Eq. \eqref{eq:B}. In general this procedure is too complicated to be insightful. In the next section we employ the above formalism to the important case of $f(R) = R+\al R^n$ in which the above equations can be solved analytically. Before that, it is useful to check our results with those of  CDL in GR.  

\subsection{CDL solution}
For GR, \(C_3=C_4=C_5=0\), so the solution becomes
\be
x=\bar{\rho}^2=\frac{-C_2}{C_1}\,,
\ee
which, after a straightforward calculation,  results in the expression for the radius of the bubble in 
the CDL tunneling from dS to dS
\ba
\rho_{\text{cdl}}^2 &=&\frac{1}{\frac{1}{4}\Big(\frac{1}{R_s}+\frac{R_s}{\Lambda_1^2}+\frac{R_s}{\Lambda_2^2}\Big)^2-\frac{R_s^2}{\Lambda_1^2\Lambda_2^2}} \nonumber \\
&=&\frac{1}{\frac{1}{\Lambda_1^2}+\Big(\frac{\epsilon}{3\sigma}-\frac{\sigma}{4M_P^2}\Big)^2}\,, 
\ea
where we have defined \(R_s \equiv\frac{2M_P^2}{\sigma}\). 

If the true vacuum is a Minkowski space i.e.  $U_2=0$ and 
\(\Lambda_2\to\infty\),  then
\be
\label{rhoCDL-eq}
\rho_{\text{cdl}}=\frac{2\Lambda_1^2/R_s}{1+\Lambda_1^2/R_s^2}\\
=\frac{\rho_0}{\gamma}\,,    \qquad  (U_2=0) , 
\ee
where \(\rho_0 \equiv2\Lambda_1^2/R_s=3\sigma/\epsilon\) is the bubble radius in the absence of gravity and \(\gamma \equiv 1+\left(\frac{\rho_0}{2\Lambda_1}\right)^2\) is the prefactor due to GR. Note that in the the current case where $U_2=0$, 
 \(\Lambda_1^2=\frac{U_1}{3M_P^2}=\frac{\epsilon}{3M_P^2}\).  
 
 Having obtained the extremum value of the bubble radius with  
$\bar \rho=\rho_{\text{cdl}}$, the decay exponent $B$ is obtained to be 
\ba
\label{BCDL-eq}
B_{\text{cdl}}=\frac{B_0}{\gamma^2 }= \frac{2 \pi^2}{\gamma^2} (\gamma-1) \rho_0^2  \, ,
\ea
where $B_0$ is the value of the tunneling exponent in flat spacetime given by Eq. (\ref{B0-eq}). 

It should be noted that the type A bounce in GR is valid if the following condition is met \cite{Weinberg:2012pjx}
\be
\epsilon>\frac{3\sigma^2}{4 M_P^2}\,,
\ee
which is translated to the condition that  \(\gamma<2\) in the above formulas.

\section{Example: \(f(R)=R+\al R^n\)}
\label{sec:example}
In this section we employ the formalism developed  above to theories with \(f(R)=R+\al R^n\) in which $\alpha$ is a constant of dimension $M^{2(1-n)}$. Models with  \(\al>0,\:n>1\) can be used to obtain inflation. For example, the Starobinsky model corresponds to $n=2$ with 
$ \al \sim (3\times10^{-5}M_P)^{-2}$ in order to be consistent with the observations. 
In the regime \(\al<0,\:0<n<1\), it can be used for dark energy models. In this section, however,  we restrict ourselves to models  with $n>1$ as in inflationary models.  

Our assumption is that the correction to GR from $f(R)$ is subleading so the theory  to leading order is given by the conventional GR. This is motivated from the fact that the corrections to GR becomes important in high energy limit so at lower energies one should recover the standard GR. In our model with  \(f(R)=R+\al R^n\) this approximation 
corresponds to 
\be
\label{approximation}
\al R^{n-1}\ll1\, .
\ee

The first step is to solve the de Sitter solution of the theory, namely solving Eq. \eqref{eq:desric} which results in
\be
R-\al(n-2)R^n=\frac{12}{\Lambda^2}\,.
\ee
The exact solution to this equations is not easy. However, employing the approximation Eq. (\ref{approximation}) we can solve it perturbatively in \(\al\). The solution to leading order in  \(\al\) is 
\be
L=\Lambda \Big[1-\frac{\al(n-2)}{2}\Big(\frac{12}{\Lambda^2}\Big)^{n-1}\Big]+\order{\al^2}\,.
\ee
Curiously we note that for \(n=2\) the solution is the same as GR, i.e.  \(L=\Lambda\). 

Inside the wall we have to solve Eq. \eqref{eq:wall} which is 
\be
\label{R-eq}
R(\xi) = R_{(0)}+\al\left[(n-2)R_{(0)}^n+3n(n-1)\left(R_{(0)}^{n-2} R'_{(0)}\right)'\right]\,.
\ee
Here we have defined  
\be
R_{(0)}(\xi) \equiv \frac{1}{M_P^2}(6U(\phi)-2U_1)\, ,
\ee
which is the value of $R(\xi)$ in the GR limit when $\alpha=0$ so the terms in the bracket in Eq. (\ref{R-eq}) represent the correction 
in $R(\xi) $ to leading order in $\alpha$.  In addition, in obtaining the above result we have approximated the d'Alembertian with the second derivative inside the wall in the limit of 
thin wall approximation.

The next step is to compute \(S_1\) and \(S_2\). For \(S_1\) from Eq. (\ref{Bw})  we have
\be
S_1=\sigma+\al(n-1)M_P^2\int\dd{\xi}\left(R_{(0)}^n-R_{(0)1}^n\right)\,,
\ee
where the subscript \(1\), as always, means the value of the corresponding variable at the 
false vacuum value. Here we have used the fact that the Ricci scalar at either sides of the wall is constant so its derivative is zero. We can write this in more compact form defining the parameter $z$ via
\be
z\equiv\frac{n-1}{(M_P^2)^{n-1}}\frac{\int\dd{\xi}\big[ (6U-2U_1)^n-(4U_1)^n\big] }{\int\dd{\xi}2(U-U_1)}\,,
\ee
so that 
\be
S_1=\sigma(1+\al z)\,.
\ee
Similarly, one can show that
\be
S_2=\al y \sigma\,,
\ee
with the new parameter $y$ defined as 
\be
y \equiv \frac{n}{(M_P^2)^{n-2}}\frac{\int\dd{\xi}\big[ (6U-2U_1)^{n-1}-(4U_1)^{n-1}\big] }{\int\dd{\xi}2(U-U_1)}\,.
\ee
The parameters $z$ and $y$ depend on $n$ and the potential $U(\phi)$. In particular, for the Starobinsky model with $n=2$, we have $y=6$ regardless of the potential while \(z\) must be computed or each potential.  In order to get some insights on their values and their dependence on the model parameters, we calculate them for a specific potential later on. 

We can compute \(\bar{\rho}\) to first order in \(\alpha\) from Eq. \eqref{eq:Bderiv}. Since the expression is too complicated in its general form, here we present our analytical results for the  case of tunneling from a de Sitter vacuum to a Minkowski vacuum. In this case, we obtain  
\be
\label{eq:rhogen}
\bar{\rho}=\rho_{\text{cdl}}\left\{1+\al\left[\frac{48^{n-1}(n\gamma^2-(6n-4)\gamma+6n-4)(\gamma-1)^{n-2}}{2\gamma(\rho_0^2)^{n-1}}+\frac{2y(2-\gamma)\gamma}{\rho_0^2}+\frac{z(2-\gamma)}{\gamma}\right]\right\}\,,
\ee
where $\rho_{\text{cdl}}$ is the size of the bubble in CDL analysis given in Eq. (\ref{rhoCDL-eq}). Correspondingly, the tunneling exponent is obtained to be 
\be
\label{eq:bgen}
B=B_{\text{cdl}}\left\{1+\al\left[\frac{48^{n-1}(2\gamma^2-(3n-2)\gamma+6n-4)(\gamma-1)^{n-2}}{\gamma(\rho_0^2)^{n-1}}-\frac{24y\gamma}{\rho_0^2}+\frac{4z}{\gamma}\right]\right\}\,,
\ee
in which $B_{\text{cdl}}$ is the decay exponent in CDL analysis given in Eq. (\ref{BCDL-eq}). In particular, for Starobinsky theory with \(n=2\), one obtain simpler expressions
\be
\bar{\rho}=\rho_{\text{cdl}}\Big[ 1+\frac{\al(2-\gamma)}{\gamma}\Big(\frac{12(\gamma^2-4\gamma+8)}{\rho_0^2}+z\Big)\Big]\,  \quad \quad (n=2) 
\ee
and
\be
\label{eq:bn2}
B=B_{\text{cdl}}\Big[1-\frac{\al}{\gamma}\Big( \frac{48(\gamma^2+4\gamma-8)}{\rho_0^2}- 4z\Big)\Big]\,, 
\quad \quad (n=2) 
\ee
where we have used the fact that \(y=6\) in this case. We see clearly that  the size of the bubble at the time of nucleation and the tunneling exponent $B$ depend non-trivially 
on $\alpha$ (calculated to first order) and $n$. 

We may deduce some general conclusions from Eq. \eqref{eq:rhogen}. Note that \(z\) and \(y\) are always positive. Also we have \(1\leq\gamma\leq2\) so the sign of the expression in the brackets may be deduced from  the sign of \(n\gamma^2-(6n-4)\gamma+6n-4\). For \(1<n\leq2\), which includes Starobinsky's theory, it is always nonnegative so we have \(\bar{\rho}>\rho_{\text{cdl}}\) e.g. in these theories the radius of the bubble at the moment of formation is bigger than the CDL bubble. It is worth noting that the radius of the bubble in GR is smaller than in the flat spacetime. As a result, higher order curvature terms compensate the effects of conventional gravity.

For \(n>2\), similar deduction holds in the interval
\be
1\leq\gamma\leq\frac{\sqrt{3n-2}}{n}\left(\sqrt{3n-2}-\sqrt{n-2}\right)<2\,.
\ee
However, for greater \(\gamma\) it should be checked numerically whether  or not the bubble radius is bigger than the CDL bubble.

Unfortunately, similar general conclusions can not be stated for the tunneling exponent 
Eq. \eqref{eq:bgen} since the relative sign of the last two terms is negative and it is not trivial to compare their magnitudes. However, for the case  \(n=2\) from Eq. \eqref{eq:bn2} one can show that 
 in the interval $ 1\leq\gamma\leq2(\sqrt{3}-1)\approx1.4\,, $ \(B>B_{\text{cdl}}\)
while for greater \(\gamma\), nothing general can be deduced.

It is interesting to note that one can tune the potential so that the bubbles have bigger sizes compared to CDL bubbles  while at the same time the tunneling rate is higher.  This is because the domain of  the parameter space where  \(B>B_{\text{cdl}}\)  does not  match in general with the domain where \(\bar{\rho}>\rho_{\text{cdl}}\). This is rather counterintuitive since  for both cases of flat spacetime and GR, bigger bubbles have lower rate (greater decay exponent). However, we see that this is not true in general in $f(R)$ theories.  

\begin{figure}[t]
	\centering
	\includegraphics[scale=0.6]{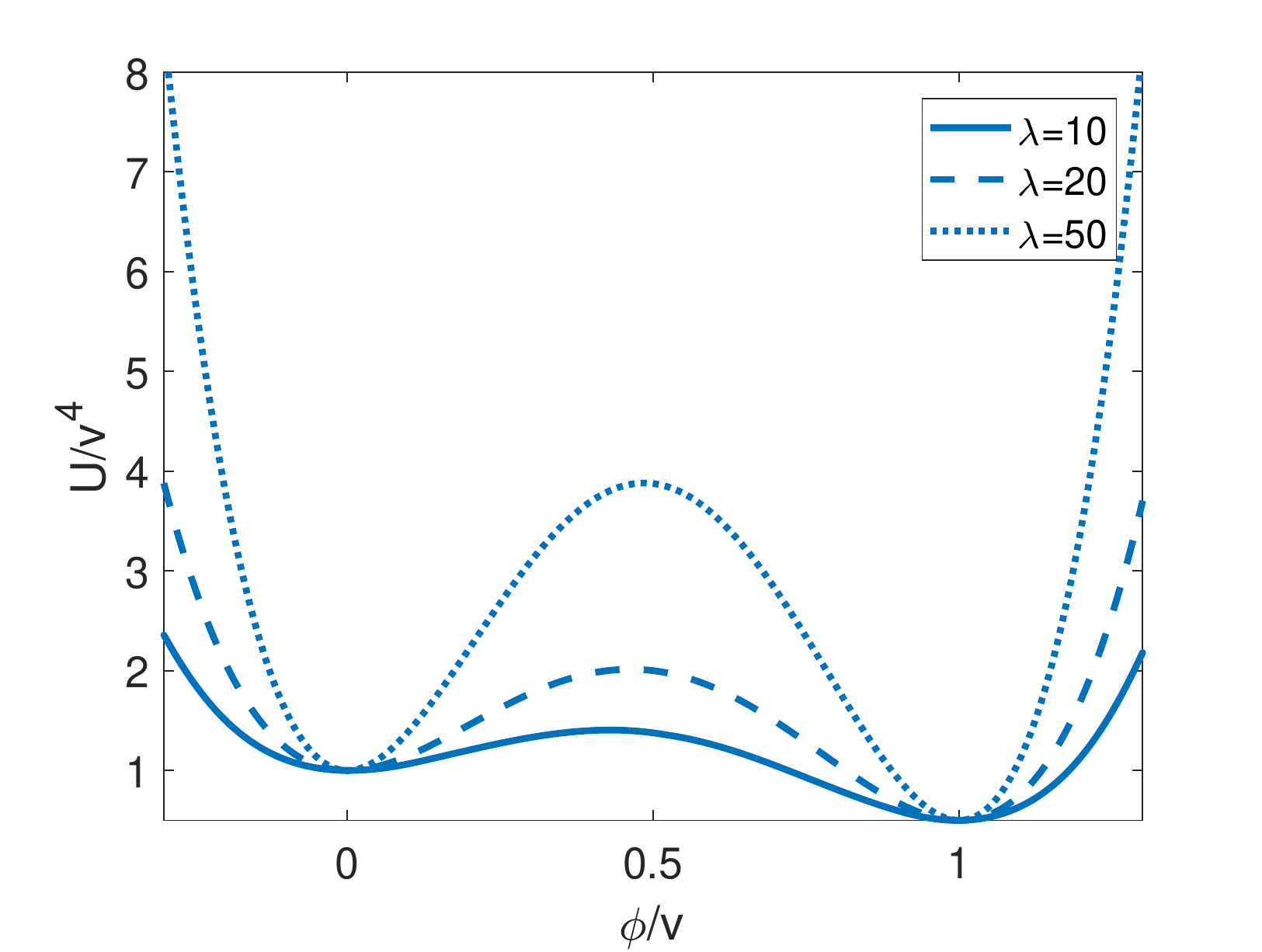}
	\caption{Potential in Eq. \eqref{eq:numpotential} with \(\lambda=10,\, 20\) and \(50\) for \(U_1=v^4\) and \(\epsilon=0.5v^4\)}
	\label{fig:numpot}
\end{figure}

So far we did not specify the form of the potential. However,  in order to get some insights about the numerical values of the various physical parameters, similar to \cite{Gregory:2018bdt,Masoumi:2012yy}, we consider the following potential 
\be
\label{eq:numpotential}
U(\phi)=\lambda\phi^4-2\lambda\left(v-\frac{\epsilon}{\lambda v^3}\right)\phi^3+\lambda\left(v^2-\frac{3\epsilon}{\lambda v^2}\right)\phi^2+U_1\,.
\ee
This is the most general form of the potential till quartic order.   Without loss of generality we have fixed the false vacuum at \(\phi_1=0\) so there is no linear term present. True vacuum occurs at \(\phi_2=v\) with the condition \(\lambda v^4>3\epsilon\). Parameters \(\epsilon\) and \(U_1\) have their understood meaning and \(\lambda\) is a dimensionless scaling. The advantage of this form compared to \cite{Masoumi:2012yy} is that interesting quantities like \(\epsilon\) can be varied without affecting others. Fig. \ref{fig:numpot} shows the  potential for different values of  \(\lambda\). It is evident that by increasing \(\lambda\), the top of the barrier is increased while other important properties do not change.

\begin{figure}[t!]
	\centering
	\includegraphics[scale=0.5]{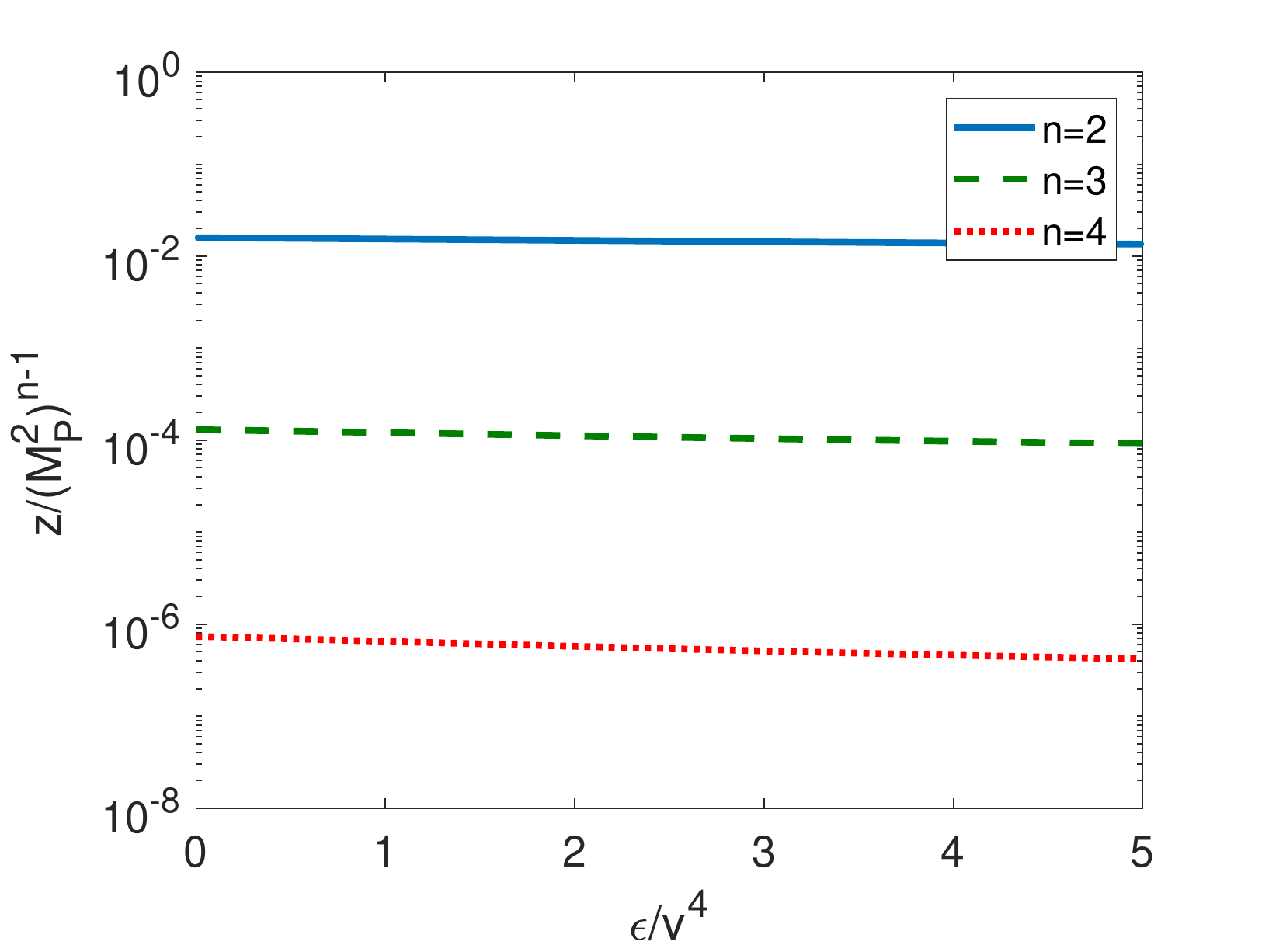}
	\includegraphics[scale=0.5]{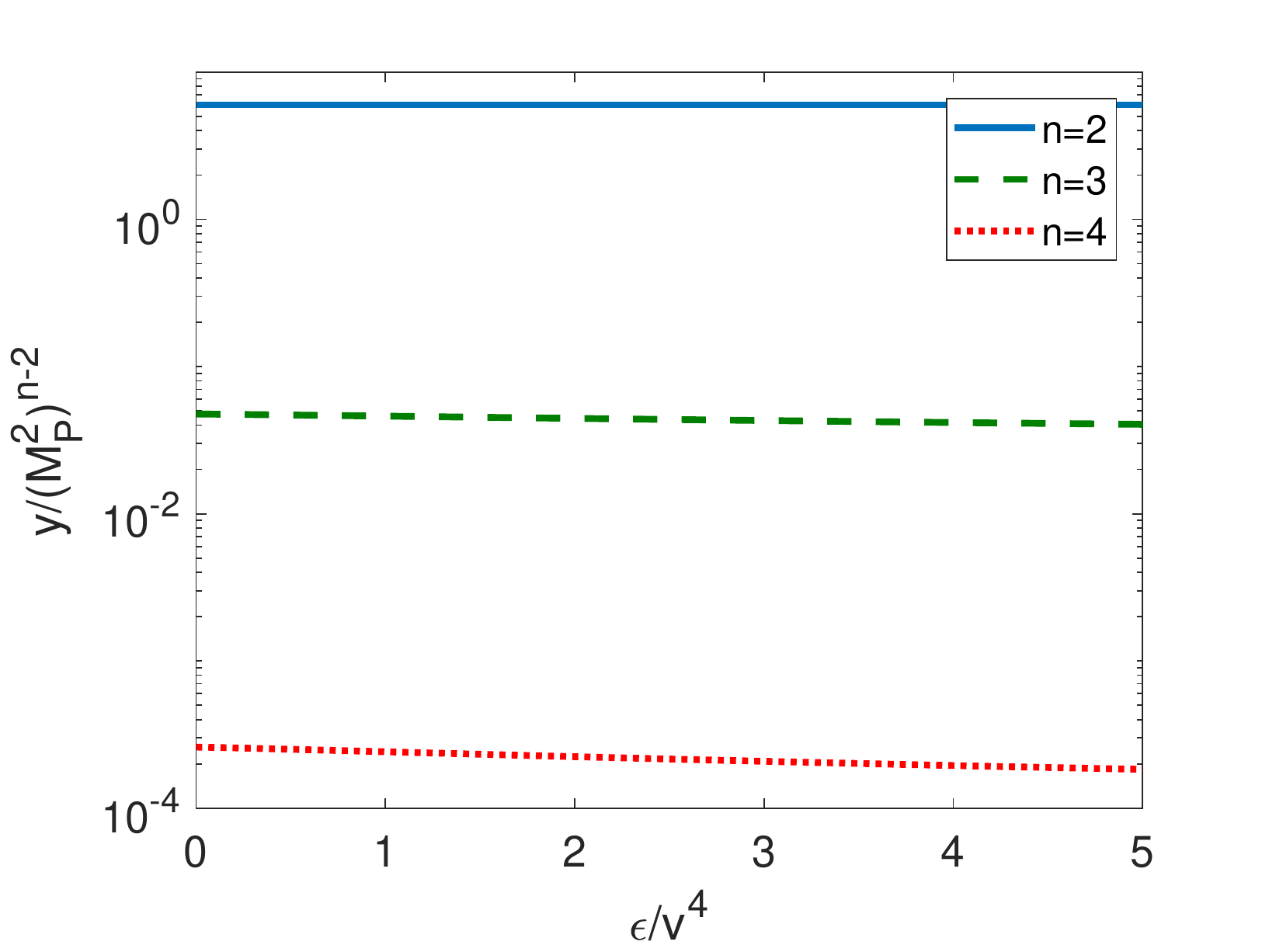}
	\caption{ The parameters \(z\) (left) and  \(y\) (right) as  functions of  \(\epsilon\) for \(n=2\), 3 and 4 for the potential Eq. \eqref{eq:numpotential}.   In both cases \(\lambda=50\) and \(U_1=5v^4\).}
	\label{fig:zyepsilon}
\end{figure}

\begin{figure}[t!]
	\centering
	\includegraphics[scale=0.5]{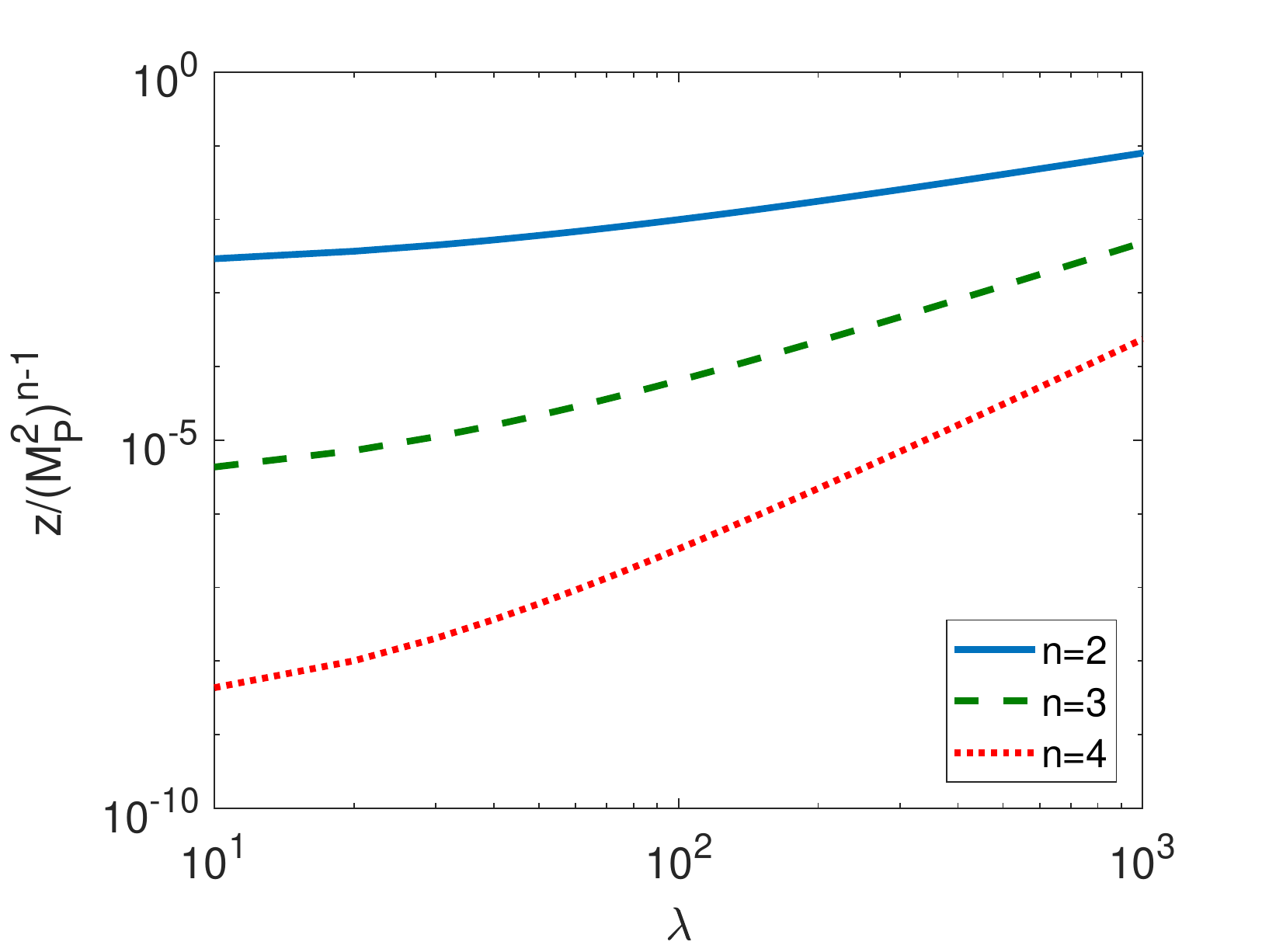}
	\includegraphics[scale=0.5]{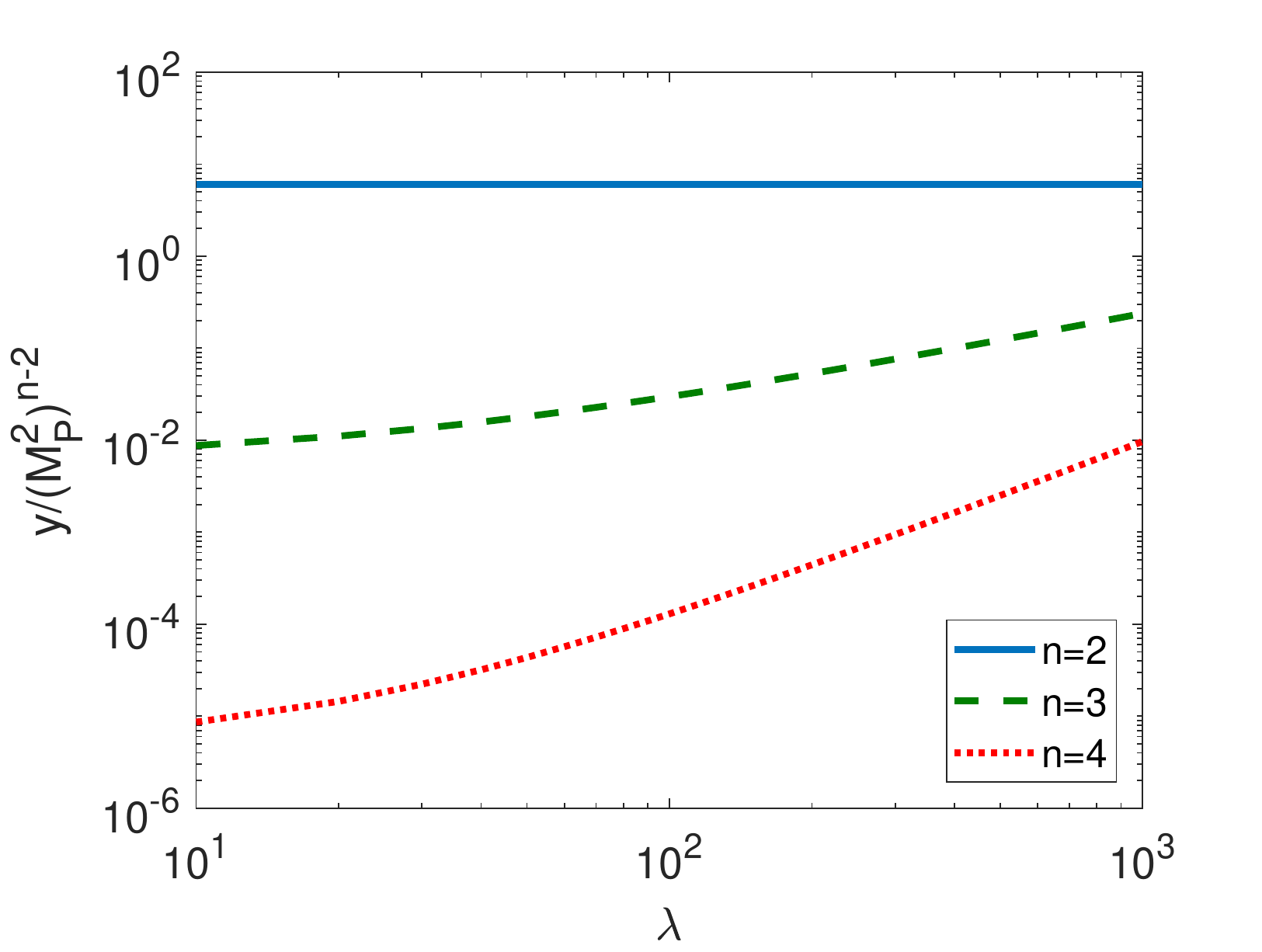}
	\caption{The parameters \(z\) (left) and  \(y\) (right) as  functions of \(\lambda\) for \(n=2\), 3 and 4 for the potential Eq. \eqref{eq:numpotential}. In both cases \(\epsilon=0.1U_1=0.5v^4\).}
	\label{fig:zylambda}
\end{figure}

For this potential, the  parameters \(z\) and \(y\) can be computed for different values of \(\epsilon\) and \(\lambda\) which are shown in Fig. \ref{fig:zyepsilon} and \ref{fig:zylambda}. Other parameters are specified in the captions of the figures. In addition,  we have set \(h\equiv\frac{v}{M_P}=0.1\). It can be seen that both parameters have negligible dependence on \(\epsilon\) while they normally increase by increasing  \(\lambda\).

\begin{figure}[t!]
	\centering
	\includegraphics[scale=0.5]{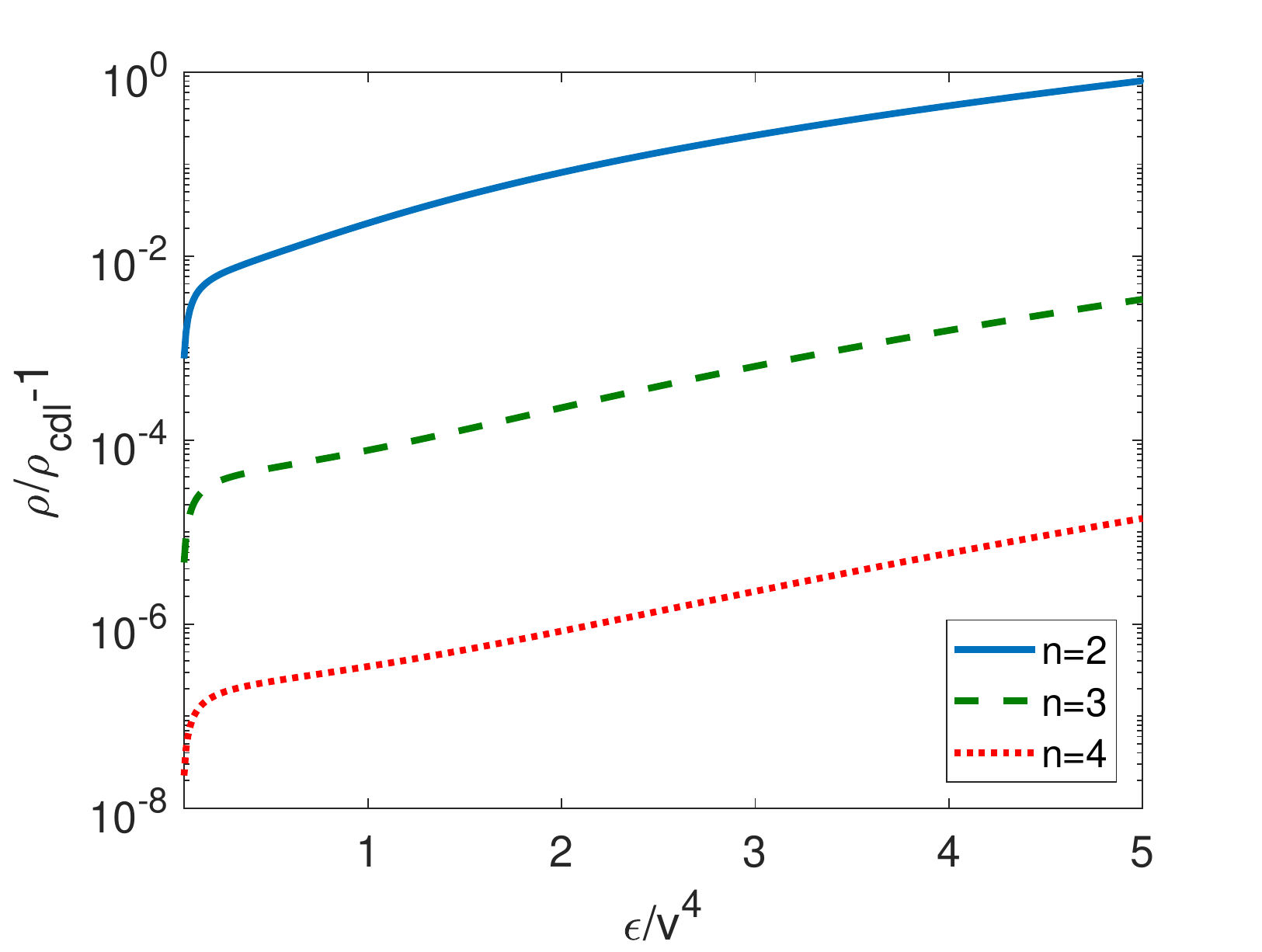}
	\includegraphics[scale=0.5]{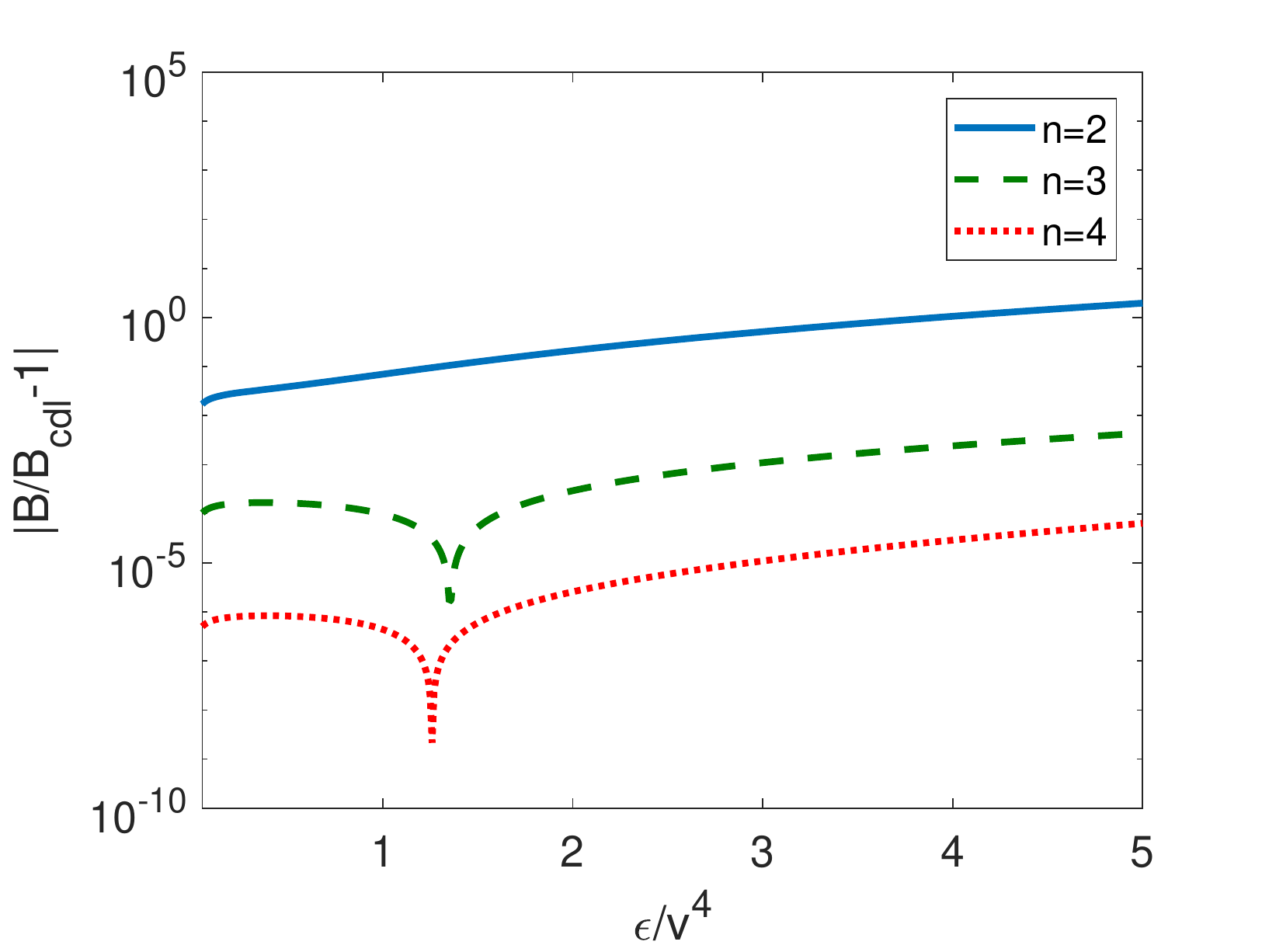}
	\caption{ The quantities \(\left(\bar{\rho}/\rho_{\text{cdl}}-1\right)\) (left) and  \(\big|B/B_{\text{cdl}}-1\big|\) (right) as functions of $\epsilon$ for cases \(n=2\), 3 and 4. For all cases, \(\lambda=100\),  \(h=0.1\) and the  vertical axis are in log scales. In the right panel $\left(B/B_{\text{cdl}}-1\right)$ vanishes at \(\epsilon\approx 1.26v^4\) for $n=3, 4$, generating the apparent singularities.}
\label{fig:rhoBe}
\end{figure}

The dependence  of the bubble radius and the decay exponent  to  \(\epsilon\) are shown in Fig. \ref{fig:rhoBe} where the quantities \(\left(\bar{\rho}/\rho_{\text{cdl}}-1\right)\) and \(\big|B/B_{\text{cdl}}-1\big|\) are plotted with  \(\lambda=100\) and \(h=0.1\). In order to have a better visualization, the vertical axes are in log scales. The apparent singularities for the decay exponent (right panels) are because the corresponding quantity vanishes at \(\epsilon\approx 1.26v^4\). The range of parameters are such that \(\gamma\) covers the whole interval $ 1 \leq  \gamma \leq 2$.  We have set \(\alpha=1\) in appropriate units of $M_{P}$ since $\alpha$ scales everything equally to first order. 

\begin{figure}[t!]
	\centering
	\includegraphics[scale=0.5]{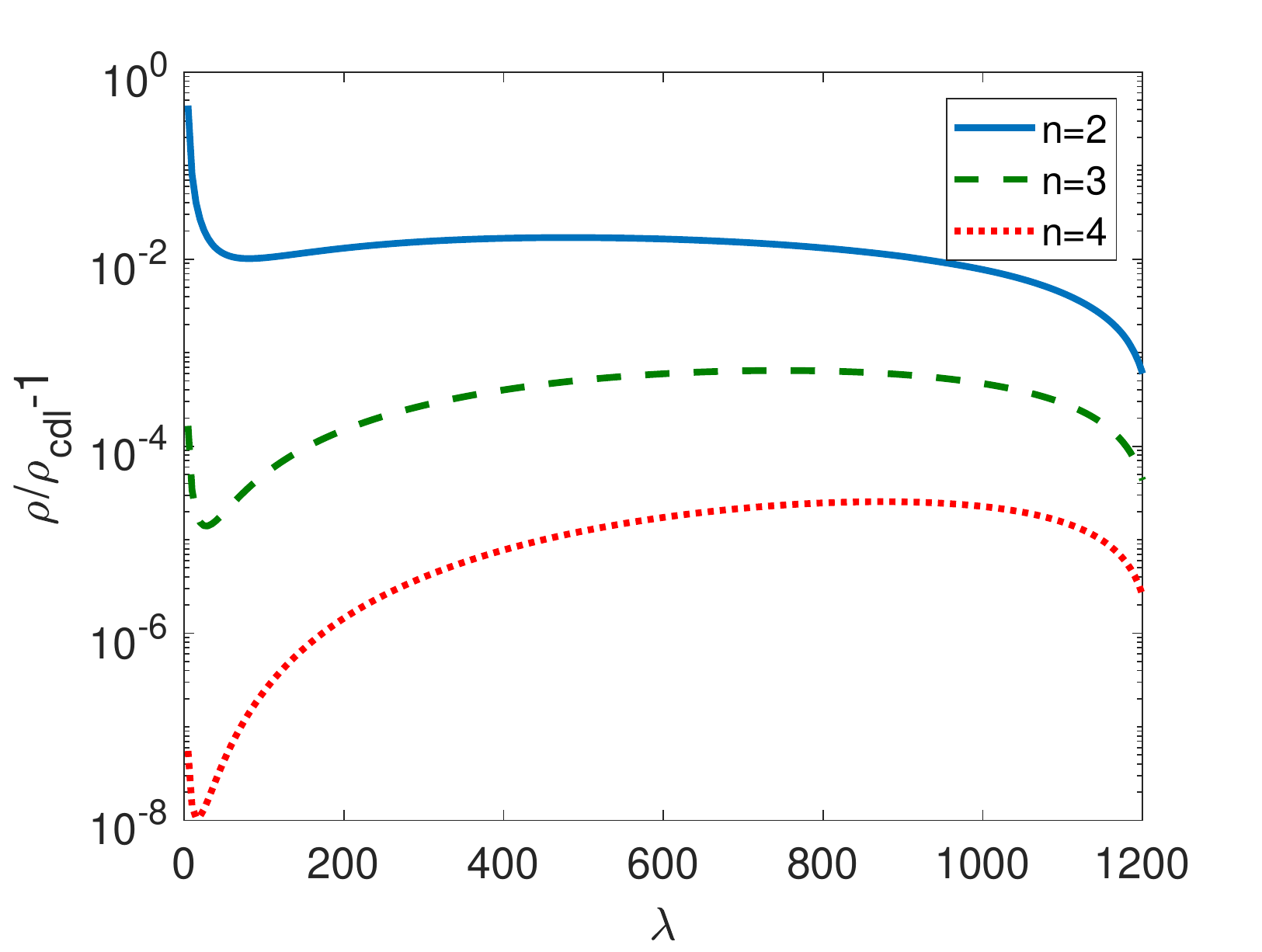}
	\includegraphics[scale=0.5]{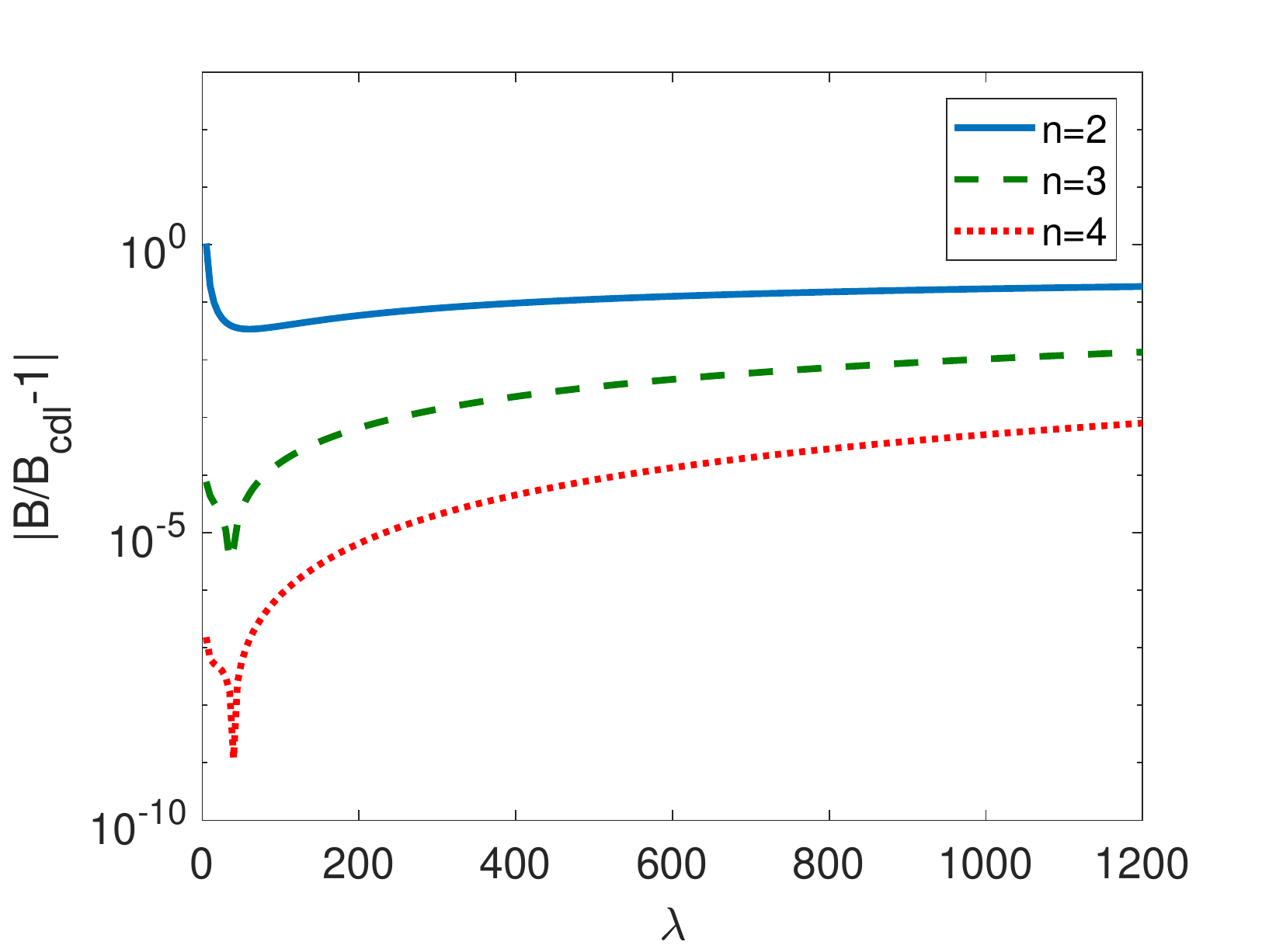}
	\caption{ The quantities \(\left(\bar{\rho}/\rho_{\text{cdl}}-1\right)\) (left) and  \(\big|B/B_{\text{cdl}}-1\big|\) (right) as functions of $\lambda$ for cases \(n=2\), 3 and 4. In all cases, \(\epsilon=0.5v^4\), \(h=0.1\) and the  vertical axis are in log scales. In the right panel $\left(B/B_{\text{cdl}}-1\right)$ vanishes at \(\lambda\approx40\) for $n=3, 4$, generating the apparent singularities.}
\label{fig:rhoBlambda}
\end{figure}

The same quantities are shown in Fig. \ref{fig:rhoBlambda}, now varying \(\lambda\) while keeping  $\epsilon$ fixed at  \(\epsilon=0.5v^4\). Now, we have the apparent singularities at \(\lambda\approx40\) where \(\big|B/B_{\text{cdl}}-1\big|\) crosses zero. 
 
It is evident that the radius of the bubble at the moment of materialization is enhanced due to the modification of gravity in our specific model. Specially, increasing \(\epsilon\) while holding $\lambda$ fixed result in a larger bubble radius.  However, the dependence of the bubble radius on \(\lambda\) is different for different values of \(n\). For Starobinsky theory, \(n=2\), it is monotonically decreasing, while for \(n=3\) and 4 it has more non-trivial dependence.

The behavior of the decay exponent is more complicated. For \(n=2\), it is bigger than the case of GR and monotonically increases by increasing \(\epsilon\). For \(n=3\) and 4, it becomes smaller than GR for \(\epsilon\gtrsim 1.26v^4\) where  \(\left(B/B_{\text{cdl}}-1\right)\) becomes negative in Fig. \ref{fig:rhoBe}. 
This is a manifestation of the fact that in these theories bubbles with bigger sizes and yet with a higher rate than in GR  can be formed. Similarly, decay exponent as a function of \(\lambda\), drops below that of GR for 
\(\lambda\gtrsim40\).

\section{Einstein frame formulation}
\label{sec:einstein}

The formalism developed above were based on Jordan frame. However, it is well-known that the $f(R) $ higher curvature gravity is equivalent to a theory containing  the standard Einstein gravity coupled minimally  to a scalar field. Intuitively speaking, the new scalar field  degree of freedom plays the role of the non-trivial function $F(R) \neq 1$. 
With this discussion in mind, one may ask how our previous result in Jordan frame can be translated in Einstein frame. One expects that the physical results to remain invariant under the change of frames, it is only the interpretation which changes. In this section we redo our analysis of tunneling in Einstein frame and verify that indeed the physical results are the same as in Jordan frame.  
 
The action in the Einstein frame with the metric $\hat{g}_{\mu\nu}$ is given by 
\be
S=\int\dd[4]{x}\sqrt{-\hat{g}}\left\{\frac{1}{2}M_P^2\hat{R}-\frac{1}{2}\hat{g}^{\mu\nu}\partial_\mu\psi\partial_\nu\psi-V(\psi)\right\}+\int\dd[4]{x}\mathcal{L}_m(F(\psi)^{-1}\hat{g}_{\mu\nu},\phi)\,,
\ee 
where
\be
\label{eq:fRcoup}
F(A(\psi))=\exp(\sqrt{\frac{2}{3}}\frac{\psi}{M_P})\,,
\ee
and \(F\) is the derivative of \(f\), the original $f(R)$ theory. The factor \(\frac{2}{3}\) is just a consequence of the spacetime dimension. This is the conformal factor between the two frames i.e. \(\hat{g}_{\mu\nu}=Fg_{\mu\nu}\). This equation must be solved to find \(A(\psi)\) and then the potential is given by
\be
\label{eq:potE}
V(\psi)=\frac{1}{2}M_P^2\frac{A(\psi)F(A(\psi))-f(A(\psi))}{F(A(\psi))^2}\,.
\ee  
For GR, \(V(\psi)=0\) and \(\psi\) becomes just a constant and the action reduces to the Einstein-Hilbert action coupled minimally to matter.  

For \(f(R)=R+\al R^n\) it is
\be
V(\psi)=\frac{n-1}{2}\frac{M_P^2}{\sqrt[n-1]{\al n^n}}\frac{1}{F(\psi)^2}\sqrt[n-1]{(F(\psi)-1)^n}\,,
\ee
where \(F(\psi)\) is given by \eqref{eq:fRcoup}. 


We are interested in the Euclidean action with scalar field as matter
\be
\label{eq:SEeins}
S_E=\int\dd[4]{x}\sqrt{g}\left\{-\frac{1}{2}M_P^2R+\frac{1}{2}g^{\mu\nu}\partial_\mu\psi\partial_\nu\psi+V(\psi)+\frac{1}{2F}g^{\mu\nu}\partial_\mu\phi\partial_\nu\phi+\frac{U(\phi)}{F^2}\right\}\,,
\ee	
where we have dropped the hat over the metric for convenience.  Note that there is a coupling between \(\psi\) and \(\phi\) via the exponential function \eqref{eq:fRcoup} which is universal for any \(f(R)\) theory. 


Assuming \(O(4)\) symmetry and conformal transformation, the line element is now given by
\be
\label{eq:Eds2}
\dd{s}^2=F\dd{\xi}^2+F\rho(\xi)^2\dd{\Omega}_3^2\,,
\ee 
which is conformally related  to \eqref{eq:ds2}. 
Correspondingly,  equations of motion  for the scalar fields are 
\be
\phi''+\frac{3\rho'}{\rho}\phi'=\dv{U}{\phi}\,,
\ee
and
\be
\psi''+\left(\frac{F'}{F}+\frac{3\rho'}{\rho}\right)\psi'=F\dv{V}{\psi}-\frac{\phi'^2+4U}{\sqrt{6}M_PF}\,.
\ee
The first one is the familiar equation of motion for the scalar field with no change as expected. The second one is nothing but \eqref{eq:tr} if we change the variable from \(F\) to \(\psi\), validating the well-known result that the new scalar field degree of freedom  is related to the derivative of \(f(R)\). 

As before, we can eliminate the kinetic terms of \(\phi\) and \(\psi\) from \eqref{eq:SEeins}, and after doing the integration by part to eliminate the second derivative of \(\rho\) and \(F\), we obtain
\be
\label{SE-Ein}
S_E=4\pi^2M_P^2\int\dd{\xi}\left[\rho^3\frac{F^2V(\psi)+U(\phi)}{M_P^2}-3\rho F\right]\,.
\ee
Note that although in the Einstein frame we have two scalar fields in presence of Einstein gravity, the Euclidean action is different since the coefficient of \(\rho\) term is \(3F\) rather than only \(3\). The reason is that the field \(\psi\) appears directly in the metric and this results in nontrivial effects.

Using Eq.  \eqref{eq:potE}, we see that the action (\ref{SE-Ein}) is exactly the same as the action  \eqref{eq:se_final}, only expressed in terms of a different variable. The rest of the analysis goes as in Jordan frame.

\section{Summary and discussions}
\label{sec:con}
In this paper, we studied the vacuum decay for a scalar field in  \(f(R)\) gravity. The higher curvature terms 
are expected to appear in the effective theories of gravity in high energy physics. In addition, many theories of high energy physics predict various effective potential with multiple vacua. Therefore, the question of tunneling and vacuum bubble nucleation  in theories of higher curvature gravity is  well motivated. Among the theories of higher curvature gravity, $f(R)$ theories are the best candidates  which are free of the pathologies such as the ghost instabilities associated with higher curvature theories. In addition, they can be a viable model of inflation such as the Starobinsky model. 

Our formalism was general,  valid for any theory of $f(R)$ gravity. However, to present compact analytical formula, we have considered the special class of $f(R) = R + \alpha R^{n}$ with $\alpha$ as a perturbative parameter.  Our motivation is that  the higher curvature terms should appear as the high energy corrections to the Einstein gravity so we expect $\alpha R^{n-1} \ll1$. We have calculated analytically the size of the nucleated bubble and the decay exponent $B$ in the case of tunneling from a dS space to a Minkowski background. In particular, for the Starobinsky model with $n=2$,  we have seen that the 
nucleated bubble has a bigger size compared to bubble created in GR.  In addition, numerical calculations show that the rate of its materialization is lower. This is obtained for a typical potential for a range of parameters. Although we have not confirmed this analytically, it is expected to be true for potentials similar to the example studied here.  For other values of $n$, depending on the model parameters, the bubble is bigger or smaller than the CDL bubble and also the decay rate is different. An interesting phenomenon which we have observed is that a  bubble with a size bigger than the CDL bubble  but with a higher nucleation rate can be formed.  This is a counterintuitive phenomenon which happens as a  consequence of the higher curvature corrections. Intuitively one may associate the effects of higher curvature terms as changing the effective Newton constant. However, our results show that the effects of higher curvature terms are more non-trivial than simply rescaling the effective Newton constant. 

Before we close this paper, there are a number of comments worth discussing. 

\begin{itemize}

\item It is believed that a dS  spacetime has a temperature related to its curvature scale $L$ via 
\be
T=\frac{1}{2\pi L}\,.
\ee
Since in $f(R)$ theory \(L\) is in general different than \(\Lambda\), then the associated temperature is modified  in comparison to GR. For instance, for \(f(R)=R+\al R^n\) it is given by
\be
\label{eq:temp}
T=T_0\left[1+\al\frac{n-2}{2}(48\pi^2T_0^2)^{n-1}\right]\,,
\ee
to first order in \(\al\) with \(T_0=1/2\pi\Lambda\). In \cite{Brown:2007sd} the tunneling rate using thermal fluctuations in the static patch in the fixed background approximation has been calculated. Furthermore, a plausible interpretation of CDL tunneling specially for big bubbles has been presented. We believe that their thermal interpretation can be applied to our case  with the modified temperature Eq. \eqref{eq:temp}.

\item So far we have concentrated only on the CDL instanton. One may wonder the implications of our analysis to the 
Hawking-Moss instanton \cite{Hawking:1981fz}.  This is important since when the potential is too flat then the CDL bounce does not exist. In this case the decay exponent can be computed easily, yielding  
\be
B=S_E(\text{top})-E_E(\text{false vacuum})\\
=-8\pi^2M_P^2(F_{\text{top}}L_{\text{top}}^2-F_1L_1^2)\,. 
\ee
For   \(f(R)=R+\al R^n\) theory it yields 
\be
B=8\pi^2M_P^2\left((\Lambda_1^2-\Lambda_{\text{top}}^2)+2\al \times12^{n-1}\left[\frac{1}{(\Lambda_1^2)^{n-2}}-\frac{1}{(\Lambda_{\text{top}}^2)^{n-2}}\right]\right)\,,
\ee
in which $L_{\text{top}}$ and $\Lambda_{\text{top}}$ represent the scales (as defined in section \ref{sec:fR}) calculated at the top of the potential. 

Curiously we note that the correction term vanishes for Starobinsky model with $n=2$. The reason is that the Hawking-Moss solution may be viewed to have thermal origin.  For $n=2$, we have \(L=\Lambda\) (as we have observed before) so the associated de Sitter temperature does not change and the nucleation rate remains unchanged. 

\item A bounce solution is a viable solution if it has one and only one negative mode i.e.  the second variation of the Euclidean action has only one negative eigenvalue \cite{Callan:1977pt}. This has been investigated for GR in \cite{Tanaka:1992zw,Lee:2014uza, Koehn:2015hga, Khvedelidze:2000cp, Gratton:2000fj} and also for the case of non minimal coupling to gravity and vacuum decay seeded by black holes (inhomogeneities) \cite{Gregory:2013hja, Burda:2015isa, Burda:2015yfa} in  
\cite{Gregory:2018bdt}. Similar calculations must be done for \(f(R)\) gravity to check the viability of our results.

\item In this work we have studied the tunneling procedure working in the Euclidean spacetime and left out the problem of evolution of the bubble after nucleation in the Lorentzian spacetime. It is normally assumed that the Lorentzian evolution is given by the analytic continuation of the Euclidean solution. Although this procedure may not be straightforward  in curved spacetime (see fro example \cite{Visser:2017atf}), in analogy to the flat spacetime, the line element after nucleation is taken to be 
\be
\dd{s}^2=-\rho^2\dd{\tau}^2+\dd{\xi}^2+\rho^2\cosh{\tau}^2\dd{\Omega}^2\,,
\ee
where we have replaced \(\chi\to i\tau\). Another approach to this problem is to solve directly the equations of motion for the nucleated bubble in the presence of gravity. In GR this is done via Israel junction conditions  \cite{Blau:1987,Berezin:1987bc}. The junction conditions for \(f(R)\) theories of gravity are developed in \cite{Deruelle:2007pt,Senovilla:2013vra}. One important difference is that the Ricci scalar and the trace of the extrinsic curvature must be continuous over the junction surface. This is in contradiction with the fact that inside and outside of the bubble are in different vacua. The  evolution of the bubble after nucleation in $f(R)$ theories is an open question which we leave for future studies.

\item There is another formalism to derive equations of motion from the action of a gravitational theory, namely the Palatini formalism. In this approach one treats the metric and the connection coefficients as independent fields. For the 
Einstein-Hilbert action this makes no difference compared to the conventional approach and it turns out that the connection coefficients must be the Levi-Civita connection. However, for \(f(R)\) theories the equations of motion are modified.  Specifically, it can be shown that in the Einstein frame the Euclidean action is 
\be
S_E=\int\dd[4]{x}\sqrt{\Tilde{g}}\left[-\frac{1}{2}M_P^2\Tilde{R}+V(\psi)+\frac{1}{2\psi}\Tilde{g}^{\mu\nu}\partial_\mu\phi\partial_\nu\phi+\frac{U(\phi)}{\psi^2}\right]\,,
\ee
which is the same as \eqref{eq:SEeins} replacing \(F\) with \(\psi\) everywhere. The important difference is that there is no kinetic term for the \(\psi\) field i.e. \(f(R)\) theories in Palatini formalism are Brans-Dicke theories with \(\omega_{BD}=-\frac{3}{2}\) \cite{Sotiriou:2008rp} so that the field \(\psi\) is non-dynamical. The equation of motion for \(\psi\), assuming \(SO(4)\)  symmetry,  is
\be
\psi^3\dv{V}{\psi}=\frac{1}{2}\phi'^2+2U(\phi)\,.
\ee
This will change the solutions inside the wall of the bubble in the thin wall limit and will affect the final result.

\end{itemize}

\vspace{0.5cm}


{\bf Acknowledgments:}  We  thank Ruth Gregory, Nima Khosravi, Sadra Jazayeri, Mehrdad Mirbabayi, Soroush Shakeri   and Takahiro Tanaka for insightful discussions and comments.  We also thank the Yukawa Institute for Theoretical Physics at Kyoto University for hospitality during the YITP symposium YKIS2018a ``General Relativity -- The Next Generation --".  B. S. thanks ICTP for hospitality during the progress of this work. H. F. thanks ICG for hospitality where this work was in its final stage.




	
	
	
	


\begin{thebibliography}{99}

\bibitem{Sher:1988mj} 
M.~Sher, 
Phys.\ Rept.\  {\bf 179}, 273 (1989).

\bibitem{Belavin:1975fg} 
A.~A.~Belavin, A.~M.~Polyakov, A.~S.~Schwartz and Y.~S.~Tyupkin, 
Phys.\ Lett.\ B {\bf 59}, 85 (1975).

\bibitem{Guth:1980zm} 
A.~H.~Guth, 
Phys.\ Rev.\ D {\bf 23}, 347 (1981), [Adv.\ Ser.\ Astrophys.\ Cosmol.\  {\bf 3}, 139 (1987)].

\bibitem{Guth:1982pn} 
A.~H.~Guth and E.~J.~Weinberg, 
 Nucl.\ Phys.\ B {\bf 212}, 321 (1983).
 
\bibitem{Coleman:1977py} 
S.~R.~Coleman, 
Phys.\ Rev.\ D {\bf 15}, 2929 (1977) Erratum: [Phys.\ Rev.\ D {\bf 16}, 1248 (1977)].

\bibitem{Callan:1977pt} 
C.~G.~Callan, Jr. and S.~R.~Coleman, 
Phys.\ Rev.\ D {\bf 16}, 1762 (1977).

\bibitem{Coleman:1980aw} 
S.~R.~Coleman and F.~De Luccia, 
Phys.\ Rev.\ D {\bf 21}, 3305 (1980).

\bibitem{Parke:1982pm} 
S.~J.~Parke, 
Phys.\ Lett.\  {\bf 121B}, 313 (1983).

\bibitem{Starobinsky:1980te} 
A.~A.~Starobinsky, 
Phys.\ Lett.\ B {\bf 91}, 99 (1980).

\bibitem{Akrami:2018odb} 
Y.~Akrami {\it et al.} [Planck Collaboration], 
[arXiv:1807.06211 [astro-ph.CO]].
  
\bibitem{Ade:2015lrj} 
P.~A.~R.~Ade {\it et al.} [Planck Collaboration], 
Astron.\ Astrophys.\  {\bf 594}, A20 (2016), [arXiv:1502.02114 [astro-ph.CO]].

\bibitem{Sotiriou:2008rp}
T.~P.~Sotiriou and V.~Faraoni, 
Rev.\ Mod.\ Phys.\  {\bf 82}, 451 (2010) [arXiv:0805.1726 [gr-qc]].

\bibitem{DeFelice:2010aj}
A.~De Felice and S.~Tsujikawa, 
Living Rev.\ Rel.\  {\bf 13}, 3 (2010) [arXiv:1002.4928 [gr-qc]].

\bibitem{Cai:2008ht} 
R.~G.~Cai, B.~Hu and S.~Koh, 
Phys.\ Lett.\ B {\bf 671}, 181 (2009) [arXiv:0806.2508 [hep-th]].

\bibitem{Charmousis:2008ce}
C.~Charmousis and A.~Padilla, 
JHEP {\bf 0812}, 038 (2008) [arXiv:0807.2864 [hep-th]].

\bibitem{Czerwinska:2016fky} 
O.~Czerwińska, Z.~Lalak, M.~Lewicki and P.~Olszewski, 
JHEP {\bf 1610}, 004 (2016) [arXiv:1606.07808 [hep-ph]].

\bibitem{Labrana:2018bkw}
P.~Labrana and H.~Cossio, 
[arXiv:1808.09291 [gr-qc]].

\bibitem{Weinberg:2012pjx} 
E.~J.~Weinberg, {Classical solutions in quantum field theory : Solitons and Instantons in High Energy Physics}. 

\bibitem{Coleman:1985rnk} 
  S.~Coleman, ``Aspects of Symmetry : Selected Erice Lectures''.

\bibitem{Banks:1973ps} 
T.~Banks, C.~M.~Bender and T.~T.~Wu, 
Phys.\ Rev.\ D {\bf 8}, 3346 (1973).

\bibitem{Tanaka:1993ez} 
T.~Tanaka, M.~Sasaki and K.~Yamamoto, 
Phys.\ Rev.\ D {\bf 49}, 1039 (1994).

\bibitem{Bitar:1978vx} 
K.~M.~Bitar and S.~J.~Chang, 
Phys.\ Rev.\ D {\bf 18}, 435 (1978).

\bibitem{Coleman:1977th}
S.~R.~Coleman, V.~Glaser and A.~Martin, 
Commun.\ Math.\ Phys.\  {\bf 58}, 211 (1978).

\bibitem{Marvel:2007pr} 
K.~Marvel and N.~Turok,
[arXiv:0712.2719 [hep-th]].

\bibitem{Dyer:2008hb}
E.~Dyer and K.~Hinterbichler, 
Phys.\ Rev.\ D {\bf 79}, 024028 (2009) [arXiv:0809.4033 [gr-qc]].


\bibitem{Visser:2017atf}
M.~Visser, 
[arXiv:1702.05572 [gr-qc]].

\bibitem{Blau:1987}
S.~Blau, E.~I.~Guendelman and A.~Guth, 
Phys. \ Rev. \ D {\bf 35}, 1747 (1987).

\bibitem{Berezin:1987bc}
V.~A.~Berezin, V.~A.~Kuzmin and I.~I.~Tkachev, 
Phys.\ Rev.\ D {\bf 36}, 2919 (1987).

\bibitem{Deruelle:2007pt}
N.~Deruelle, M.~Sasaki and Y.~Sendouda, 
Prog.\ Theor.\ Phys.\  {\bf 119}, 237 (2008) [arXiv:0711.1150 [gr-qc]].

\bibitem{Senovilla:2013vra}
J.~M.~M.~Senovilla, 
Phys.\ Rev.\ D {\bf 88}, 064015 (2013) [arXiv:1303.1408 [gr-qc]].


\bibitem{Gregory:2018bdt}
R.~Gregory, K.~M.~Marshall, F.~Michel and I.~G.~Moss, 
[arXiv:1808.02305 [hep-th]].

\bibitem{Masoumi:2012yy} 
A.~Masoumi and E.~J.~Weinberg, 
Phys.\ Rev.\ D {\bf 86}, 104029 (2012) [arXiv:1207.3717 [hep-th]].

\bibitem{Brown:2007sd}
A.~R.~Brown and E.~J.~Weinberg, 
Phys.\ Rev.\ D {\bf 76}, 064003 (2007) [arXiv:0706.1573 [hep-th]].

\bibitem{Hawking:1981fz} 
  S.~W.~Hawking and I.~G.~Moss, 
  Phys.\ Lett.\  {\bf 110B}, 35 (1982), 
  [Adv.\ Ser.\ Astrophys.\ Cosmol.\  {\bf 3}, 154 (1987)].

\bibitem{Tanaka:1992zw}
T.~Tanaka and M.~Sasaki, 
Prog.\ Theor.\ Phys.\  {\bf 88}, 503 (1992).

\bibitem{Lee:2014uza}
H.~Lee and E.~J.~Weinberg, 
Phys.\ Rev.\ D {\bf 90}, no. 12, 124002 (2014) [arXiv:1408.6547 [hep-th]]. 

\bibitem{Koehn:2015hga} 
  M.~Koehn, G.~Lavrelashvili and J.~L.~Lehners,
  Phys.\ Rev.\ D {\bf 92}, no. 2, 023506 (2015), 
  [arXiv:1504.04334 [hep-th]].

\bibitem{Khvedelidze:2000cp} 
  A.~Khvedelidze, G.~V.~Lavrelashvili and T.~Tanaka,
  Phys.\ Rev.\ D {\bf 62}, 083501 (2000), 
  [gr-qc/0001041].
  
\bibitem{Gratton:2000fj} 
  S.~Gratton and N.~Turok,
  Phys.\ Rev.\ D {\bf 63}, 123514 (2001), 
  [hep-th/0008235].


\bibitem{Gregory:2013hja} 
  R.~Gregory, I.~G.~Moss and B.~Withers,
  JHEP {\bf 1403}, 081 (2014), 
  [arXiv:1401.0017 [hep-th]].

\bibitem{Burda:2015isa} 
  P.~Burda, R.~Gregory and I.~Moss, 
  Phys.\ Rev.\ Lett.\  {\bf 115}, 071303 (2015), 
  [arXiv:1501.04937 [hep-th]].

\bibitem{Burda:2015yfa} 
  P.~Burda, R.~Gregory and I.~Moss, 
  JHEP {\bf 1508}, 114 (2015), 
  [arXiv:1503.07331 [hep-th]].



	
\end{thebibliography}
\end{document}